\documentclass[aps,prl,superscriptaddress,floatfix,a4paper,twocolumn,superscriptaddress,noshowpacs]{revtex4-1}
\usepackage{graphicx}
\usepackage{amsmath}
\usepackage{amssymb}
\usepackage{color}
\usepackage{float}

\usepackage[colorlinks=true,citecolor=blue,linkcolor=blue]{hyperref}

\begin{document}

\flushbottom

\title{Coherent coupling between vortex bound states and magnetic impurities in 2D layered superconductors}

\author{Sunghun Park}
\affiliation{Departamento de F\'isica Te\'orica de la Materia Condensada, Instituto Nicol\'as Cabrera and Condensed Matter Physics Center (IFIMAC), Universidad Aut\'onoma de Madrid, E-28049 Madrid,
Spain}

\author{V\'ictor Barrena}
\affiliation{Laboratorio de Bajas Temperaturas y Altos Campos Magn\'eticos, Departamento de F\'isica de la Materia Condensada, Instituto Nicol\'as Cabrera and Condensed Matter Physics Center (IFIMAC), Unidad Asociada UAM-CSIC, Universidad Aut\'onoma de Madrid, E-28049 Madrid,
Spain}

\author{Samuel Ma\~nas-Valero} 
\affiliation{Instituto de Ciencia Molecular (ICMol), Universidad de Valencia, Catedr\'atico Jos\'e Beltr\'an 2, 46980 Paterna, Spain}

\author{Jos\'e J. Baldov\'i}
\affiliation{Instituto de Ciencia Molecular (ICMol), Universidad de Valencia, Catedr\'atico Jos\'e Beltr\'an 2, 46980 Paterna, Spain}
\affiliation{Max Planck Institute for the Structure and Dynamics of Matter, Luruper Chaussee 149, D-22761 Hamburg, Germany}

\author{Ant\'on Fente}
\affiliation{Laboratorio de Bajas Temperaturas y Altos Campos Magn\'eticos, Departamento de F\'isica de la Materia Condensada, Instituto Nicol\'as Cabrera and Condensed Matter Physics Center (IFIMAC), Unidad Asociada UAM-CSIC, Universidad Aut\'onoma de Madrid, E-28049 Madrid,
Spain}

\author{Edwin Herrera}
\affiliation{Laboratorio de Bajas Temperaturas y Altos Campos Magn\'eticos, Departamento de F\'isica de la Materia Condensada, Instituto Nicol\'as Cabrera and Condensed Matter Physics Center (IFIMAC), Unidad Asociada UAM-CSIC, Universidad Aut\'onoma de Madrid, E-28049 Madrid,
Spain}

\author{Federico Mompe\'an}
\affiliation{Instituto de Ciencia de Materiales de Madrid, Consejo Superior de Investigaciones Cient\'{\i}ficas (ICMM-CSIC), Sor Juana In\'es de la Cruz 3, 28049 Madrid, Spain}

\author{Mar Garc{\'i}a-Hern{\'a}ndez}
\affiliation{Instituto de Ciencia de Materiales de Madrid, Consejo Superior de Investigaciones Cient\'{\i}ficas (ICMM-CSIC), Sor Juana In\'es de la Cruz 3, 28049 Madrid, Spain}

\author{\'Angel Rubio}
\affiliation{Max Planck Institute for the Structure and Dynamics of Matter, Luruper Chaussee 149, D-22761 Hamburg, Germany}
\affiliation{Nano-Bio Spectroscopy Group and European Theoretical Spectroscopy Facility (ETSF), Universidad del Pa\'is Vasco CFM CSIC-UPV/EHU-MPC \& DIPC, Avenida Tolosa 72, 20018 San Sebasti\'an, Spain}

\author{Eugenio Coronado} 
\affiliation{Instituto de Ciencia Molecular (ICMol), Universidad de Valencia, Catedr\'atico Jos\'e Beltr\'an 2, 46980 Paterna, Spain}

\author{Isabel Guillam\'on}
\affiliation{Laboratorio de Bajas Temperaturas y Altos Campos Magn\'eticos, Departamento de F\'isica de la Materia Condensada, Instituto Nicol\'as Cabrera and Condensed Matter Physics Center (IFIMAC), Unidad Asociada UAM-CSIC, Universidad Aut\'onoma de Madrid, E-28049 Madrid,
Spain}

\author{Alfredo Levy Yeyati}
\affiliation{Departamento de F\'isica Te\'orica de la Materia Condensada, Instituto Nicol\'as Cabrera and Condensed Matter Physics Center (IFIMAC), Universidad Aut\'onoma de Madrid, E-28049 Madrid,
Spain}

\author{Hermann Suderow}
\affiliation{Laboratorio de Bajas Temperaturas y Altos Campos Magn\'eticos, Departamento de F\'isica de la Materia Condensada, Instituto Nicol\'as Cabrera and Condensed Matter Physics Center (IFIMAC), Unidad Asociada UAM-CSIC, Universidad Aut\'onoma de Madrid, E-28049 Madrid,
Spain}

\begin{abstract}
Bound states in superconductors are expected to exhibit a spatially resolved electron-hole asymmetry which is the hallmark of their quantum nature. This asymmetry manifests as oscillations at the Fermi wavelength, which is usually tiny and thus washed out by thermal broadening or by scattering at defects. Here we demonstrate theoretically and confirm experimentally that, when coupled to magnetic impurities, bound states in a vortex core exhibit an emergent axial electron-hole asymmetry on a much longer scale, set by the coherence length. 
We study vortices in 2H-NbSe$_2$ and in 2H-NbSe$_{1.8}$S$_{0.2}$ with magnetic impurities, characterizing these with detailed Hubbard-corrected density functional calculations. 
We find that the induced electron-hole imbalance depends on the band character of the superconducting material.  Our results open interesting prospects for the study of coupled superconducting bound states.
\end{abstract}

\maketitle


\begin{figure*}
	\begin{center}
	\centering
	\includegraphics[width=0.9\textwidth]{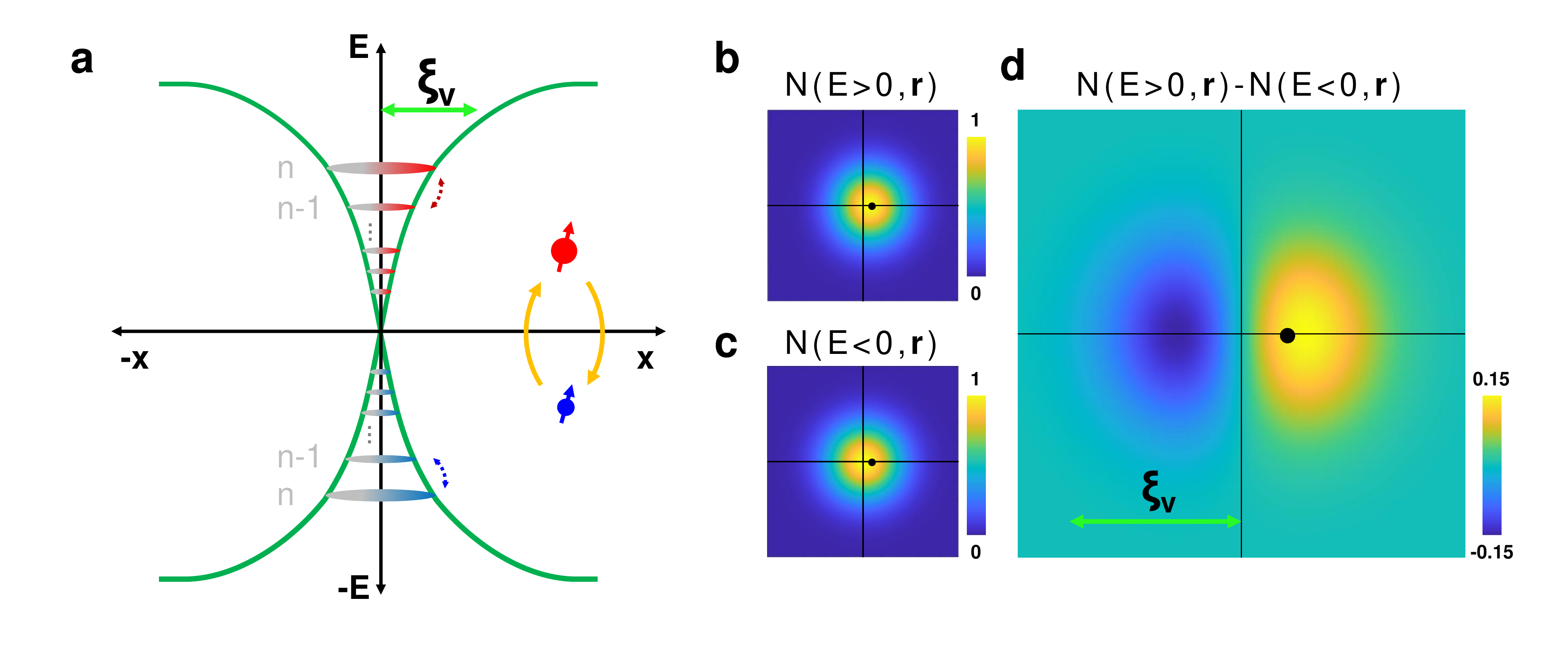}
	\end{center}
	\vskip -0.5 cm

	\caption{{\bf Interaction between CdGM and YSR states.} {\bf a} The YSR state is schematically represented by the red and blue dots with arrows representing the spin. The value of the superconducting gap as a function of the radius is shown in green for energies above and below the Fermi level. The CdGM states are schematically shown by colored circles. The difference in energy between levels, $\Delta^2/E_F$, with $\Delta$ the superconducting gap and $E_F$ the Fermi energy, is strongly exaggerated. Green arrow on top schematically represents the vortex core size $\xi_V$. The interaction between CdGM and YSR bound states leads to a spatial shift in the CdGM states, which is different for electron and hole components of the LDOS inside the superconducting gap, $N(E>0)$ ({\bf b}) and $N(E<0)$ ({\bf c}). The difference is shown in {\bf d} and the LDOS is larger along the direction of the YSR impurity, as corresponding to a hole band. For an electron band, we expect a different shift in {\bf b} and {\bf c}, such that it leads to an inversion of the color scale in {\bf d}. The YSR impurity is represented by a black dot and the vortex core center by the crossing point between black lines in {\bf b}, {\bf c} and {\bf d}.}
	\label{FigShiba}
	\end{figure*}


Bound states appear in superconductors at localized perturbations in the superconducting order parameter. Both Caroli de Gennes Matricon (CdGM) states at vortex cores \cite{Abrikosov1957,Caroli1964} and Yu-Shiba-Rusinov (YSR) states at magnetic impurities \cite{Yu1965,Shiba1968,Rusinov1969}
are examples of this phenomenon. YSR states provide mixed electron-hole excitations that serve to create the conditions needed for Majorana states. These are expected for instance at the ends of magnetic chains of YSR atoms \cite{PhysRevB.84.195442,Nadj-Perge2014,PhysRevLett.115.197204,Kimeaar5251}. On the other hand, CdGM states have been proposed to isolate and manipulate Majoranas in a topological superconductor \cite{,PhysRevLett.98.237002,doi:10.1146/annurev-conmatphys-030212-184337,Alicea_2012,PhysRevLett.100.096407,Sun2017,PhysRevB.91.214514}. 

The nature of YSR and CdGM states is, however, quite different.
YSR states are spin polarized and appear at a single or a few subgap energies and exhibit oscillations at the Fermi wavelength $\lambda_F$ that can be resolved with atomic scale local density of states (LDOS) measurements
\cite{Yazdani1997,Flatte1997,Menard2015,HEINRICH20181}.
By contrast, CdGM states are spin degenerate and form a quasi-continuum with a level separation $\Delta^2/E_F$ (where $\Delta$ is the superconducting gap and $E_F$ is the Fermi energy), which is usually small compared to $\Delta$. Thus, their discreteness and their mixed electron hole character only appears at very low temperatures or for $\Delta\approx E_F$ and in absence of scattering, in the so called quantum limit \cite{Hayashi1998}. Otherwise, thermal excitations or defects produce dephasing resulting in an electron-hole symmetric LDOS pattern at vortex cores.

Thus, in most cases, CdGM states are electron-hole symmetric and their features in the LDOS extend to much larger distances than those of YSR states. Here we ask the question if we can build a hybrid quantum system consisting in vortices close to magnetic impurities and transfer the quantum property of YSR states, i.e. their electron-hole asymmetry, into the more extended CdGM states far from the quantum limit.

%



As we show below, we indeed theoretically predict and experimentally observe electron-hole asymmetric features in the LDOS of vortices in presence of magnetic impurities. As we schematically represent in Fig.\,\ref{FigShiba}, a magnetic impurity close to a vortex core induces a coupling between CdGM states with $n$ and $n \pm 1$ angular momenta. This coupling produces a slight shift of the charge density of the positive (negative) energy excitations towards (outwards) the impurity with respect to their mean position, which remains even away of the quantum limit. The sign of the coupling changes when the bands at the Fermi energy have a hole (electron) character. The discrete nature of CdGM states is thus revealed in the vortex core LDOS: the difference between the electron and hole LDOS would exhibit an axial asymmetry as illustrated in
Fig.\,\ref{FigShiba}d, with a larger LDOS close to the position of the magnetic impurity. We emphasize that, in contrast to the above mentioned oscillations at the tiny $\lambda_F$ scale, the electron-hole asymmetric feature in the vortex LDOS occurs at a much larger length scale.

\begin{figure*}
	\includegraphics[width=0.9\textwidth]{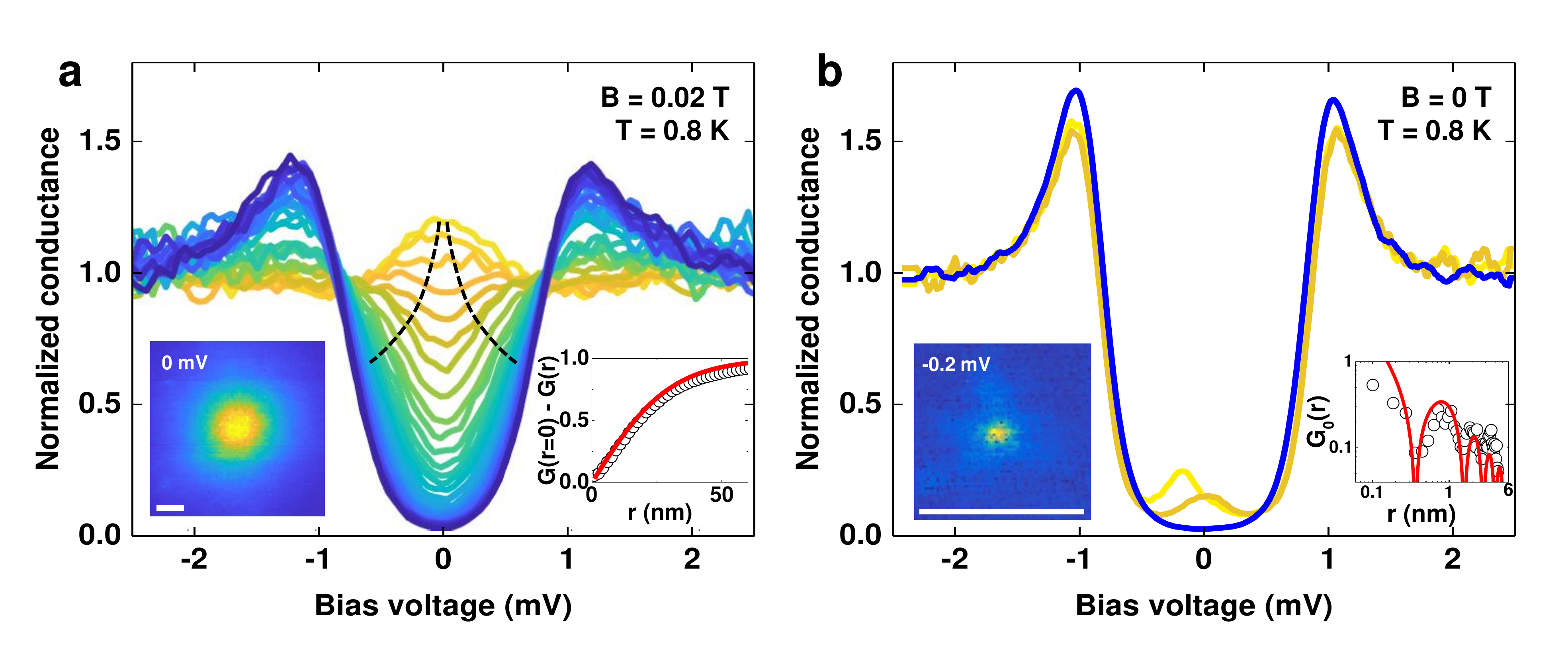}
	\caption{{\bf Length scales of CdGM and YSR states.} {\bf a} Tunneling conductance from the center of a vortex (yellow to blue) in 2H-NbSe$_{1.8}$S$_{0.2}$. Dashed lines represent schematically the splitting of the zero bias conductance peak when leaving the vortex core, a characteristic feature of CdGM states. In the lower left inset we show a zero bias conductance map. In the lower right panel we show as white circles the radially averaged zero bias conductance $G$ as a function of the radius. {\bf b} Tunneling conductance as a function of the position at a magnetic Fe impurity (yellow and ochre curves are close to the impurity center and separated approximately by $\lambda_F\approx$ 0.7 nm) and far from it (blue) in 2H-NbSe$_{1.8}$S$_{0.2}$. The tunneling conductance map at the bias voltage where we observe a maximum of the conductance in the yellow curve is shown in the lower left panel. In the lower right panel as white circles we show the spatial dependence of the conductance along a line from the position of the impurity outwards. Red lines in the lower right insets follows roughly the expected dependencies for the spatial variation of the LDOS for CdGM ({\bf a}) and YSR ({\bf b}) states, discussed in previous work \cite{Hess1989,Hess1990,Guillamon2008c,Menard2015}. White bars in the lower left insets of {\bf a}, {\bf b} are both 10 nm long.}
	\label{FigScales}
	\end{figure*}

The transition metal dichalcogenide 2D layered superconductor 2H-NbSe$_2$ is the first material where the vortex LDOS has been measured and one of the few where the nature of CdGM states has been extensively studied, both in experiment and theory \cite{Hess1989,Hess1990,Guillamon2008c}. YSR impurities have been also imaged in detail in this material \cite{Menard2015,Liebhaber2020,Kezilebieke2018}.
Vortex cores in 2H-NbSe$_2$ ($T_c=7.2$ K) are highly anisotropic, with a characteristic sixfold star shape. Previous work imaged YSR impurities and vortex cores at the same time, but did not identify any particular connection \cite{Menard2015}. On the other hand, when doped with S as in 2H-NbSe$_{1.8}$S$_{0.2}$ ($T_c=6.6$ K) the vortex core CdGM states are in-plane isotropic, leading to round shaped, symmetric, vortex cores \cite{Fente2016}. Using these two systems, we can study the interaction between YSR and CdGM states for in-plane isotropic (2H-NbSe$_{1.8}$S$_{0.2}$) and anisotropic (2H-NbSe$_2$) vortices.

To produce the YSR states in 2H-NbSe$_{2}$ and 2H-NbSe$_{1.8}$S$_{0.2}$ we introduce Fe impurities during sample growth (about 150 ppm), as identified after the experiment using inductively coupled plasma atomic analysis. In this diluted regime Fe impurities produce practically no changes in the residual resistivity or $T_c$ both for 2H-NbSe$_{1.8}$S$_{0.2}$ or 2H-NbSe$_2$. The amount of Fe impurities is sufficiently small as to leave the superconducting gap and vortex structure unaffected, but large enough to be easily detected in the area occupied by a single vortex. We use a Scanning Tunneling Microscope (STM) to measure the LDOS as a function of the position at 0.8 K. Samples are cleaved at or below liquid He, to allow for a clean atomically flat surface and the tip is prepared in-situ \cite{Suderow2011}. 

It is of interest to first analyze the different length scales associated to isolated CdGM and YSR states, particularly in the case of 2H-NbSe$_{1.8}S_{0.2}$ (Fig.\,\ref{FigScales}). As we show in Fig.\,\ref{FigScales}a, CdGM states provide a zero bias peak at the center of the vortex that decays with distance at a scale which is generally larger than the coherence length (of approximately 10 nm in 2H-NbSe$_2$ and 7 nm in 2H-NbSe$_{1.8}$S$_{0.2}$ as obtained from H$_{c2}$(T)) and is magnetic field dependent\cite{Kogan2005,Fente2016}. The vortex core size at the magnetic fields considered here is of $\xi_V \approx$ 30 nm\cite{Fente2016}. The zero bias peak splits when leaving the vortex core, as shown in previous work \cite{Hess1989,Hess1990,Guillamon2008c,Gygi1991,Clinton1992,Hayashi1998,Fente2016}. The six-fold anisotropy characteristic of vortex cores in 2H-NbSe$_2$ is washed out by the S substitutional disorder in 2H-NbSe$_{1.8}S_{0.2}$. On the other hand, a YSR state in 2H-NbSe$_{1.8}S_{0.2}$ is shown in Fig.\,\ref{FigScales}b. There is a conductance peak within the superconducting gap which changes from positive to negative bias voltage values at a scale of order of $\lambda_F$ (about 0.7 nm, see lower right inset in Fig.\,\ref{FigScales}b and Ref.\,\cite{Menard2015}). At the same time, the height of the peak decreases exponentially with distance (lower right inset of Fig.\,\ref{FigScales}b). Thus we see that the effect of YSR states is transposed two orders of magnitude in distance, from $\lambda_F$ to $\xi_V$, and leads to the electron-hole asymmetric CdGM states we discuss below.

\begin{figure}
	\includegraphics[width=0.9\columnwidth]{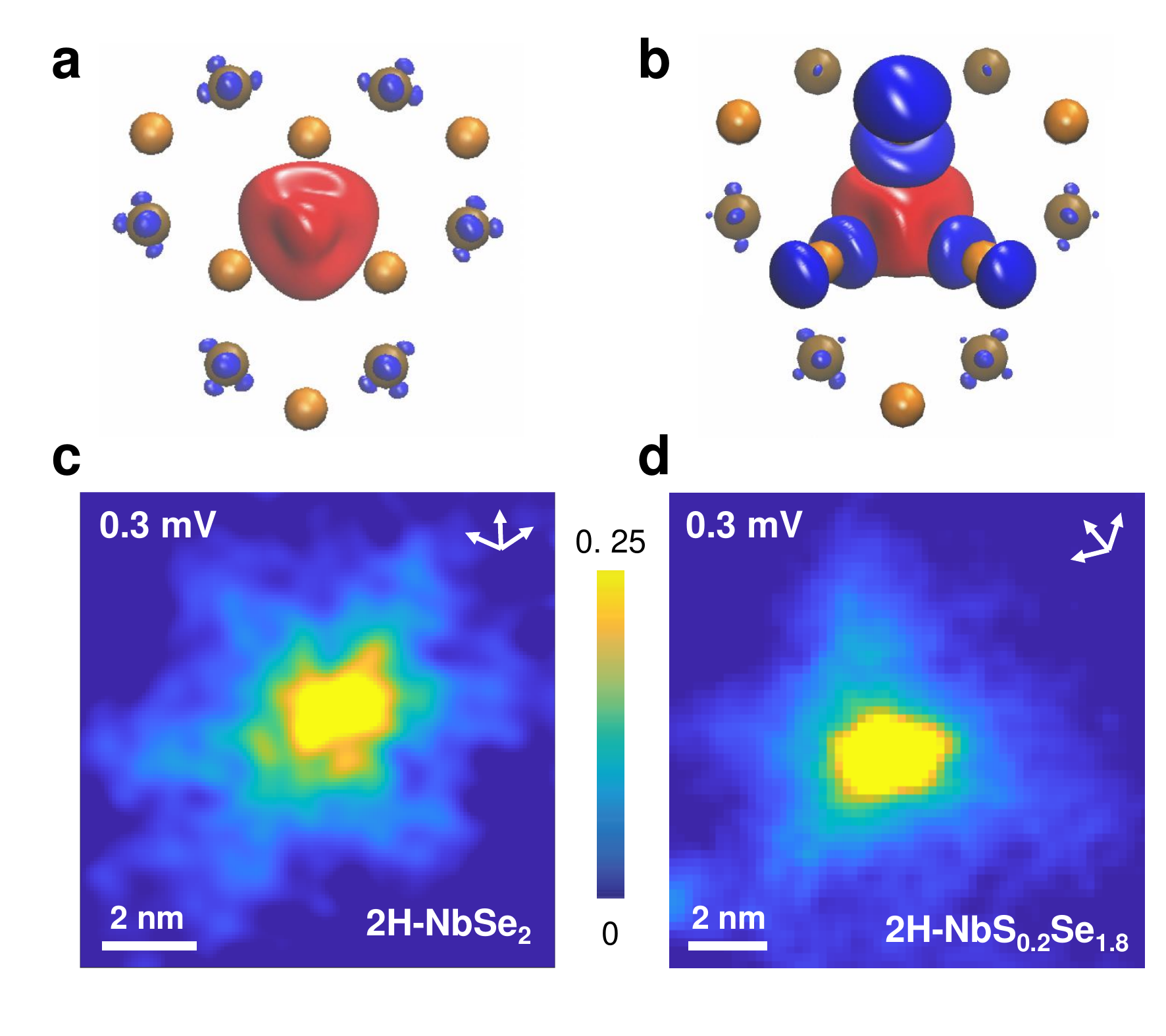}
	\caption{{\bf YSR states at magnetic impurities.} Spin-density isosurface of 2H-NbSe$_{2}$ {\bf a} and of 2H-NbSe$_{1.8}$S$_{0.2}$ {\bf b} obtained from DFT+U calculations. Red stands for spin up and blue for spin-down charge density. The isosurface is plotted for an imbalance by 0.002 of the spin density. The Fe atom (not shown) is located on the red spot, substituting a Nb atom. Se atoms are orange and Nb atoms ochre colors. We represent the two uppermost layers (Se and Nb) from the top. In {\bf b} there are S atoms in layers below, shown in the Supplementary Information. Calculations are made on four layers, using $4\times 4 \times 2$ sized slabs (see Supplementary Information for further details). {\bf c}, {\bf d} Measured tunneling conductance maps at a Fe impurity in 2H-NbSe$_2$ {\bf c} and in 2H-NbSe$_{1.8}$S$_{0.2}$ {\bf d}. White arrows provide the crystalline directions of the atomic Se lattice.}
	\label{SpinIsosurface}
	\end{figure}

To better understand the magnetism at the Fe impurities we calculated the electronic structure and the spin-density isosurface, i.e. the difference between spin-up and spin-down charge densities (Fig.\,\ref{SpinIsosurface}), by means of Hubbard-corrected density functional theory (DFT+U). We constructed a slab model with a 4 $\times$ 4 $\times$ 2 supercell that contains one isolated magnetic impurity at a Nb site. To model 2H-NbSe$_{1.8}$S$_{0.2}$, we introduced approximately 10\% S atoms (115 Se and 13 S atoms) randomly distributed. Computational methods are detailed in the Supplemental Information. In 2H-NbSe$_{2}$ we observe clearly a strong magnetic moment on the Fe atom (Fig.\,\ref{SpinIsosurface}a). The same behavior is observed at the Fe site in 2H-NbSe$_{1.8}$S$_{0.2}$ with however a large antiferromagnetic coupling with immediately neighboring Se atoms that become spin-polarized (Fig.\,\ref{SpinIsosurface}). Importantly, this coupling breaks the six-fold in-plane symmetry of the Se lattice. The introduction of S atoms lowers the symmetry from six to three fold.  We show in  Fig.\,\ref{SpinIsosurface}c,d the measured tunneling conductance map $G(\bold{r})=G(x,y)$ on a YSR state on 2H-NbSe$_2$ (Fig.\,\ref{SpinIsosurface}c) and on 2H-NbSe$_{1.8}$S$_{0.2}$ (Fig.\,\ref{SpinIsosurface}d). We see that the LDOS at YSR impurities is modified from a six-fold star shape in 2H-NbSe$_2$ to a predominantly three-fold star shape in 2H-NbSe$_{1.8}$S$_{0.2}$ and ascribe this effect to the symmetry breaking in the lattice induced by the S distribution in 2H-NbSe$_{1.8}$S$_{0.2}$, as suggested by our calculations (Fig.\,\ref{SpinIsosurface}a,b). The three fold symmetry is smeared at the scale of $\xi_V$ leading to the observed round vortex cores shown in Fig.\,\ref{FigScales}a.

\begin{figure*}
	\includegraphics[width=0.9\textwidth]{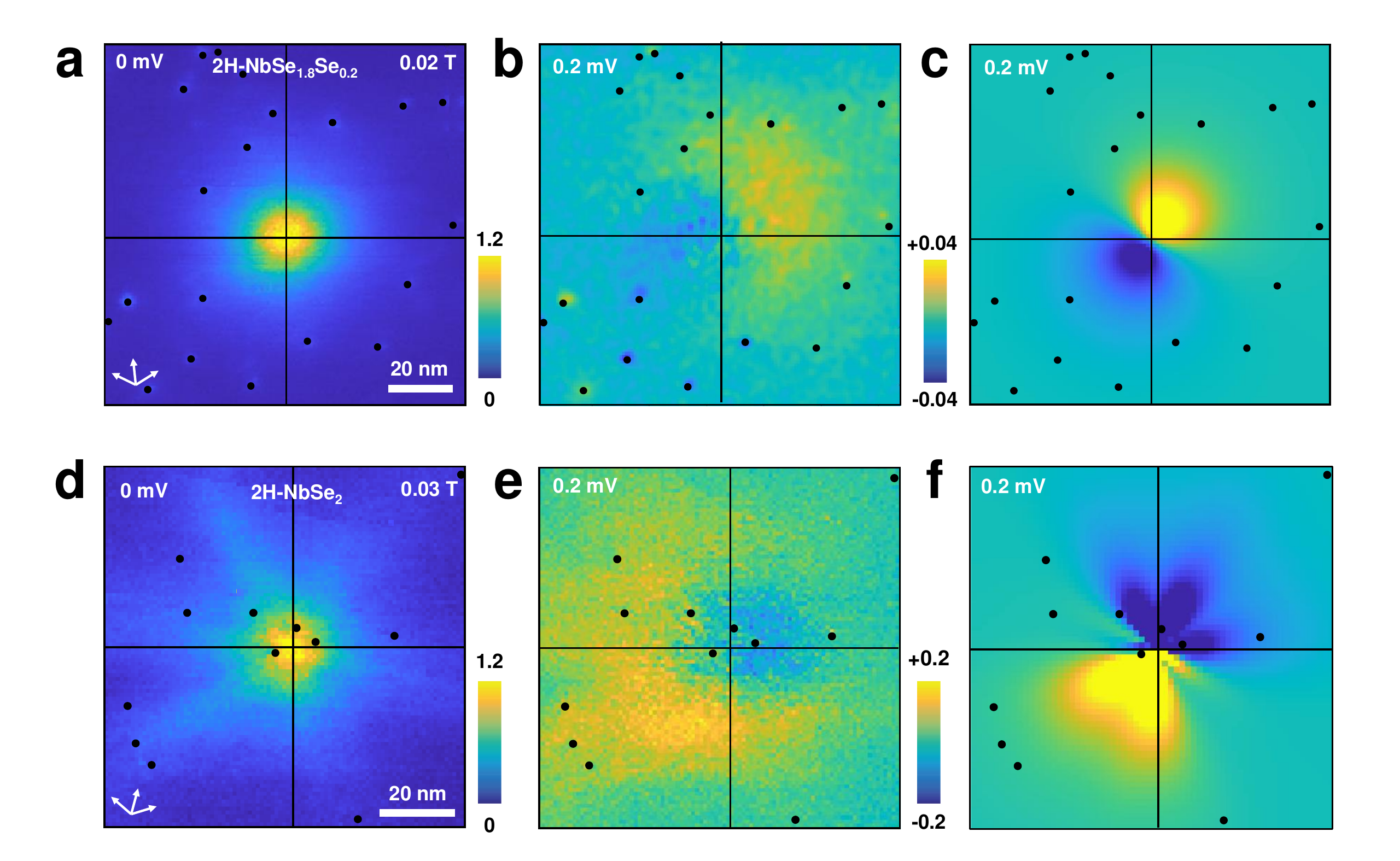}
	\caption{{\bf Vortex cores coupled with YSR states.} {\bf a} Zero bias conductance map of a vortex in 2H-NbSe$_{1.8}$S$_{0.2}$. {\bf b} Map showing the difference between the normalized tunneling conductance at positive and negative bias voltages $\frac{\delta G(\bold{r},V)}{G_0}=\frac{G(\bold{r},V)-G(\bold{r},-V)}{G_0}$ at $\vert V\vert =$0.2 mV and in the same field of view as {\bf a}. {\bf c} $\frac{\delta G(\bold{r},V)}{G_0}$ obtained from the calculation described in the text. The same quantities are plotted for 2H-NbSe$_2$ in {\bf d}, {\bf e} and {\bf f}. Color scales are given by the bars on the right side of panels {\bf a} and {\bf d} for the zero bias conductance, and in between panels {\bf b}, {\bf c} and {\bf e}, {\bf f} for $\frac{\delta G(\bold{r},V)}{G_0}$. Fe impurities are marked by black dots. Vortex centers are at the crossing point between the black lines. White arrows in {\bf a},{\bf b} give the directions of the Se lattice.}
	\label{FigVortex}
	\end{figure*}

We show vortices in close proximity to YSR impurities in Fig.\ref{FigVortex}a,d. When we make the difference between images taken at positive and negative bias voltages,  $\frac{\delta G(\bold{r},V)}{G_0} = \frac{G(\bold{r},V) - G(\bold{r},-V)}{G_0}$ (with $G_0$ the averaged tunneling conductance for bias voltages above the gap), we observe that vortex cores are not axially symmetric (Fig.\ref{FigVortex}b,e). In contrast (as we show in detail in the Supplemental Information), 
$\frac{\delta G(\bold{r},V)}{G_0}$ is axially symmetric in absence of YSR impurities.


As stated above, we trace the broken axial symmetry to the interplay between vortex and YSR states. We have calculated the perturbation to a rotationally symmetric vortex induced by a magnetic impurity located close to the vortex core. As shown in the Supplemental Information, we start with a two dimensional superconductor described by a Bogoliubov-de Gennes Hamiltonian. We find discrete energy levels $E_n$ and the shape of electron and hole wave functions $\psi_n^{+}$, $\psi_n^{-}$ of CdGM vortex bound states (with $n$ the angular momentum number). 
Magnetic YSR impurities are characterized as usual by the exchange coupling $J$ at the impurity sites \cite{Huang2020}. 
This coupling leads to an effective Hamiltonian in the subspace spanned by the states $\psi_{n-1}, \psi_{n}, \psi_{n+1}$, with solution $\tilde{\psi}_n^{+}$, $\tilde{\psi}_n^{-}$. Without YSR impurities, the vortex core LDOS obtained is always axially symmetric, as found previously. There are slight, axially symmetric electron-hole variations at the Fermi wavelength $\lambda_F$ scale, which are smeared out due to dephasing except in the quantum limit. The vortex core LDOS with YSR impurities, obtained from $\tilde{\psi}_n^{+}$, $\tilde{\psi}_n^{-}$, is however axially asymmetric. The asymmetry is due to the spatial shift in the perturbed CdGM states $\tilde{\psi}_n^{+}$, $\tilde{\psi}_n^{-}$ and is induced by the mixing between adjacent CdGM levels ($n+1$ and $n-1$). This asymmetry is roughly given by $\frac{\delta G(\bold{r},V)}{G_0}\propto |\tilde{\psi}_n^+|^2 - |\tilde{\psi}_n^-|^2 \propto \pm J^2 e^{-4 r_p/\xi_V} \cos(\theta-\theta_p)$, where $\theta$ is the polar angle with respect to the vortex center, $r_p$ and $\theta_p$ provide the length and the angle of the line joining the vortex center and the impurity position and the $\pm$ sign depends on whether the effective mass is negative or positive, i.e. whether the bands have a hole or an electron like character. The magnitude of the perturbation decays exponentially with the distance from the impurity to the vortex center.

As we show in Fig.\,\ref{FigVortex}, the observed LDOS asymmetry can be qualitatively reproduced using our theory (Fig.\,\ref{FigVortex}c,f). For that purpose we introduce an impurity distribution corresponding 
to the one in the experiments and add the contribution of each impurity to the asymmetry. Furthermore, we use an isotropic gap for 2H-NbSe$_{1.8}$S$_{0.2}$ and a six-fold anisotropic gap for 2H-NbSe$_2$. Detailed parameters of the calculation are provided in the Supplemental Information. Here we highlight that the exchange coupling $J$ is negative, corresponding to the antiferromagnetic exchange found in Fig.\,\ref{SpinIsosurface}a,b  and that we can use the same value for all the impurities. In practice, due to the already mentioned distance dependence, the asymmetry $\frac{\delta G(\bold{r},V)}{G_0}$ is however dominated by the few impurities which are closest to the vortex core.  

Let us note that vortices in 2H-NbSe$_2$, with their characteristic strong six-fold star shape, present a rather involved shape of the asymmetry (Fig.\,\ref{FigVortex}f). This suggests that the spatial extension of CdGM states determines the overall shape of the asymmetry.

In all, we conclude from our combined theoretical and experimental work that YSR states produce electron-hole asymmetric vortex cores. The YSR states allow visualizing the discrete nature of CdGM levels and their electron-hole asymmetry is translated to large scales.
Our theory also suggests that a superconductor with predominantly electron band character should lead to an opposite shift in the LDOS. For example, vortices have been observed in $\beta-$Bi$_2$Pd \cite{Herrera2015,PhysRevLett.120.167001,lv2017experimental}, which has predominantly electron character \cite{sakano2015topologically,iwaya2017full}. YSR states in $\beta-$Bi$_2$Pd have been observed \cite{PhysRevLett.120.167001} but their influence on CdGM states has not yet been addressed.

Bound states in vortex cores have been considered in the past mostly to address the influence of pair potential disturbances on vortex pinning \cite{PhysRevB.57.5457}. Recent calculations find vortex core states induced by magnetic or nonmagnetic impurities which result in small modifications of the superconducting gap parameter \cite{PhysRevB.103.024510}. However in this work we find that vortex positions and gap parameter do not exhibit visible changes at energies away from the gap edge in presence of magnetic impurities. 

Unconventional d-wave or s$\pm$ superconductors often show pair breaking at atomic impurities and vortices at the same time \cite{Fischer2007,Massee2015,Allan2013,chen2021friedel}. Our results suggest that vortex bound states might be strongly influenced by such impurities, although a theory for the interplay between YSR and CdGM states within a non s-wave pairing interaction is required to analyze this further.

Finally, let us note that one can also envisage experiments using atomic manipulation or deposition to place impurities at certain positions. The position of vortices is easily modified by changing the magnetic field. Groups of geometrically arranged YSR impurities can lead to significant spatially distortions of the LDOS of vortex states.

\acknowledgments{This work was supported by the EU (ERC-StG-679080, ERC-2015-AdG-694097, ERC AdG Mol-2D 788222 and H2020-MSCAIF-2016-751047), by EU program Cost (CA16218, Nanocohybri and MOLSPIN CA15128), by the Spanish State Research Agency (FIS2017-84330-R, RYC-2014-15093, CEX2018-000805-M, MAT2017-89993-R co-financed by FEDER and CEX2019-000919-M), by the Deutsche Forschungsgemeinschaft (DFG) under under Germany’s Excellence Strategy - Cluster of Excellence Advanced Imaging of Matter (AIM) EXC 2056 - 390715994, RTG 1995 and GRK 2247, by the Comunidad de Madrid through program NANOMAGCOST-CM (Program No.S2018/NMT-4321) and by the Generalitat Valenciana (Grupos Consolidados IT1249-19, Prometeo Programme, iDiFEDER/2018/061 and CDEIGENT/2019/022). AR acknowledges support by the MPI - 
New York City Center for Non-Equilibrium Quantum Phenomena. HS and IG acknowledge SEGAINVEX at UAM for design and construction of STM cryogenic equipment. SP acknowledges Banco Santander (ref. 102I0112).}


\end{document}


\flushbottom

\title{Supplementary Information: Coherent coupling between vortex bound states and magnetic impurities in 2D layered superconductors} 

\author{Sunghun Park}
\affiliation{Departamento de F\'isica Te\'orica de la Materia Condensada, Instituto Nicol\'as Cabrera and Condensed Matter Physics Center (IFIMAC), Universidad Aut\'onoma de Madrid, E-28049 Madrid,
Spain}

\author{V\'ictor Barrena}
\affiliation{Laboratorio de Bajas Temperaturas y Altos Campos Magn\'eticos, Departamento de F\'isica de la Materia Condensada, Instituto Nicol\'as Cabrera and Condensed Matter Physics Center (IFIMAC), Unidad Asociada UAM-CSIC, Universidad Aut\'onoma de Madrid, E-28049 Madrid,
Spain}

\author{Samuel Ma\~nas-Valero} 
\affiliation{Instituto de Ciencia Molecular (ICMol), Universidad de Valencia, Catedr\'atico Jos\'e Beltr\'an 2, 46980 Paterna, Spain}

\author{Jos\'e J. Baldov\'i}
\affiliation{Instituto de Ciencia Molecular (ICMol), Universidad de Valencia, Catedr\'atico Jos\'e Beltr\'an 2, 46980 Paterna, Spain}
\affiliation{Max Planck Institute for the Structure and Dynamics of Matter, Luruper Chaussee 149, D-22761 Hamburg, Germany}

\author{Ant\'on Fente}
\affiliation{Laboratorio de Bajas Temperaturas y Altos Campos Magn\'eticos, Departamento de F\'isica de la Materia Condensada, Instituto Nicol\'as Cabrera and Condensed Matter Physics Center (IFIMAC), Unidad Asociada UAM-CSIC, Universidad Aut\'onoma de Madrid, E-28049 Madrid,
Spain}

\author{Edwin Herrera}
\affiliation{Laboratorio de Bajas Temperaturas y Altos Campos Magn\'eticos, Departamento de F\'isica de la Materia Condensada, Instituto Nicol\'as Cabrera and Condensed Matter Physics Center (IFIMAC), Unidad Asociada UAM-CSIC, Universidad Aut\'onoma de Madrid, E-28049 Madrid,
Spain}

\author{Federico Mompe\'an}
\affiliation{Instituto de Ciencia de Materiales de Madrid, Consejo Superior de Investigaciones Cient\'{\i}ficas (ICMM-CSIC), Sor Juana In\'es de la Cruz 3, 28049 Madrid, Spain}

\author{Mar Garc{\'i}a-Hern{\'a}ndez}
\affiliation{Instituto de Ciencia de Materiales de Madrid, Consejo Superior de Investigaciones Cient\'{\i}ficas (ICMM-CSIC), Sor Juana In\'es de la Cruz 3, 28049 Madrid, Spain}

\author{\'Angel Rubio}
\affiliation{Max Planck Institute for the Structure and Dynamics of Matter, Luruper Chaussee 149, D-22761 Hamburg, Germany}
\affiliation{Nano-Bio Spectroscopy Group and European Theoretical Spectroscopy Facility (ETSF), Universidad del Pa\'is Vasco CFM CSIC-UPV/EHU-MPC \& DIPC, Avenida Tolosa 72, 20018 San Sebasti\'an, Spain}

\author{Eugenio Coronado} 
\affiliation{Instituto de Ciencia Molecular (ICMol), Universidad de Valencia, Catedr\'atico Jos\'e Beltr\'an 2, 46980 Paterna, Spain}

\author{Isabel Guillam\'on}
\affiliation{Laboratorio de Bajas Temperaturas y Altos Campos Magn\'eticos, Departamento de F\'isica de la Materia Condensada, Instituto Nicol\'as Cabrera and Condensed Matter Physics Center (IFIMAC), Unidad Asociada UAM-CSIC, Universidad Aut\'onoma de Madrid, E-28049 Madrid,
Spain}

\author{Alfredo Levy Yeyati}
\affiliation{Departamento de F\'isica Te\'orica de la Materia Condensada, Instituto Nicol\'as Cabrera and Condensed Matter Physics Center (IFIMAC), Universidad Aut\'onoma de Madrid, E-28049 Madrid,
Spain}

\author{Hermann Suderow}
\affiliation{Laboratorio de Bajas Temperaturas y Altos Campos Magn\'eticos, Departamento de F\'isica de la Materia Condensada, Instituto Nicol\'as Cabrera and Condensed Matter Physics Center (IFIMAC), Unidad Asociada UAM-CSIC, Universidad Aut\'onoma de Madrid, E-28049 Madrid,
Spain}

\maketitle

\section{S1. Bogoliubov de Gennes calculation of YSR and CdGM states}

\paragraph{Model Hamiltonian for CdGM states.} We consider the Bogoliubov de Gennes (BdG) equation describing an isolated vortex at the origin in two dimensions in the basis of electron and hole wavefunctions $\psi = (\psi^{+},\psi^{-})^{T}$,
\begin{equation}
\begin{pmatrix}
-\frac{\hbar^2 \mathbf{k}^2}{2 m^*} + {E_F} & \Delta(\mathbf{r}) \\
\Delta(\mathbf{r})^* & \frac{\hbar^2 \mathbf{k}^2}{2 m^*} -  {E_F}
\end{pmatrix} \psi(\mathbf{r}) =
E \psi(\mathbf{r}),
\end{equation}
where $m^*$ is the absolute value of the effective mass, $E_F$ is the Fermi energy and $\Delta(\mathbf{r})=\Delta_0 \text{tanh}(r/\xi_V) e^{-i \theta}$ with the size of a vortex core $\xi_V$. Notice that we consider a system with a hole like band character, as 2H-NbSe$_2$\cite{Johannes2006}. Otherwise, the diagonal components of the matrix need to be interchanged. As $\text{2H-NbSe}_2$ is a strong type II superconductor with a large penetration depth (of about 200 nm \cite{PhysRevLett.98.057003,PhysRevMaterials.4.084005}), we can take a constant magnetic field. Spin degeneracy is not included for simplicity, since the wave functions for both spin states are the same. Following previous works~\cite{Caroli1964,Bardeen1969,Clinton1992}, the CdGM bound state energy and corresponding wave function in the asymptotic region ($r \gg 1/k_F$) in the low energy limit are approximately given by 
\begin{align}
\varepsilon_n &= \frac{E_n}{\Delta_0}= (n+\frac{1}{2})  \frac{\Delta_0}{2E_F}, \label{Energy}\\
\psi_n(\mathbf{r}) &=
\begin{pmatrix}
 \psi^{+}_n(\mathbf{r})\\
 \psi^{-}_n(\mathbf{r})
\end{pmatrix} \label{VBS}\\
&= A_n(r)
\begin{pmatrix}
e^{-i (n+1)\theta} \cos(k_F r + F_n-\frac{\eta_n}{2})\\
e^{-i n \theta} \cos(k_F r + F_n+\frac{\eta_n}{2})
\end{pmatrix}. \nonumber
\end{align}
Here $A_n(r)$ is 
\begin{equation}
A_n(r) = \frac{1}{\sqrt{N_n}}\frac{e^{-\lambda_n(r)}}{\sqrt{k_F \tilde{r}_n}},
\end{equation}
where $N_n$ is a normalization factor and $\tilde{r}_{n} \equiv (r^2+((n+1/2)/k_F)^2)/r$ is introduced to avoid singularities at $r=0$. $n$ are the eigenstate numbers. $\lambda_n, F_n$, and $\eta_n$ are functions of $r$ given by
\begin{align}
F_n(r)&= \frac{\left(n+\frac{1}{2}\right)^2+\frac{1}{4}}{2 k_F \tilde{r}_n} - \frac{\pi}{2}\left(\sqrt{ \left(n+\frac{1}{2}\right)^2+\frac{1}{4}}+\frac{1}{2} \right),\\
\eta_n(r) &=\left[\left(\frac{2\alpha_n }{\xi_V} r\right)^{-1} + \left(\frac{\pi}{2} - 
|\varepsilon_n| \right)^{-1}\right]^{-1}, \\
\lambda_n(r)&=\frac{1}{2}\left(1-\frac{\varepsilon^2_n}{2}\right) 
\left[\frac{2}{\xi_V}r - \frac{1}{\beta_n} \text{tanh}\left(\frac{2\beta_n}{\xi_V} r\right) \right].
\end{align}
$\alpha_n$ and $\beta_n$ are 
\begin{align}
\alpha_n &= -\frac{\pi}{4} +\sqrt{\frac{1}{2}}\frac{1-\mu^3_n}{1-\mu^2_n},\\
\beta_n &= \sqrt{\frac{\alpha_n/2}{1-\varepsilon^2_n/2}},
\end{align}
where $\mu_n = |E_F \sqrt{2}/((n+1/2) \Delta_0)|$. The level spacing $\Delta^2_0/(2E_F)$ in Eq.~\eqref{Energy} is in agreement with the result $\approx 0.85 \Delta^2_0/(2E_F)$ from Eq. (10) in Ref.~\cite{Caroli1964} with $\Delta(\mathbf{r})=\Delta_0 \text{tanh}(r/\xi_V) e^{-i \theta}$. We set $\Delta_0 = 1$ meV, $\xi_V= 30$ nm, and $k_F = 9$ nm$^{-1}$.

The BdG equation for the Fermi level lying in an electron like band in the basis $\phi=(\phi^{+},\phi^{-})^{T}$ is 
\begin{equation}
\begin{pmatrix}
\frac{\hbar^2 \mathbf{k}^2}{2 m^*} - E_F & \Delta(\mathbf{r}) \\
\Delta(\mathbf{r})^* & -\frac{\hbar^2 \mathbf{k}^2}{2 m^*} +  E_F
\end{pmatrix} 
\begin{pmatrix}
\phi^{+}(\mathbf{r})\\
\phi^{-}(\mathbf{r})
\end{pmatrix}
=E 
\begin{pmatrix}
\phi^{+}(\mathbf{r})\\
\phi^{-}(\mathbf{r})
\end{pmatrix}.
\end{equation}
The functions $\psi$ and $\phi$ are related by a transformation, 
\begin{equation}
\begin{pmatrix}
\phi^{+}(\mathbf{r})\\
\phi^{-}(\mathbf{r})
\end{pmatrix}=\mathcal{C}
\begin{pmatrix}
0 & 1 \\
1 & 0
\end{pmatrix}
\begin{pmatrix}
\psi^{+}(\mathbf{r})\\
\psi^{-}(\mathbf{r})
\end{pmatrix}=
\begin{pmatrix}
\psi^{- *}(\mathbf{r})\\
\psi^{+ *}(\mathbf{r})
\end{pmatrix},
\end{equation}
where $\mathcal{C}$ is the complex conjugate operator. We note that the sign of phase shift $\eta_n$ between $\psi^{+}$ and $\psi^{-}$ along the radial direction shown in Eq. \eqref{VBS} is inverted to $-\eta_n$ for $\phi^{+}$ and $\phi^{-}$ when the band character changes from a hole like band to electron like band. We will show below that the difference between the electron and hole components of the LDOS depends on the sign of $\eta_n$ (see Eq. \eqref{Pftn}), leading to the dependence of the axial asymmetry on the band character.

It is useful to remember the consequences of these expressions for the shape of the LDOS at and around a vortex core\cite{Caroli1964,Bardeen1969,Clinton1992,Gygi1991,Fischer2007,Hayashi1998,PhysRevB.54.10094,PhysRevB.103.024510}. The electron and hole LDOS follows approximately the sum over all $\lvert\psi_n^+(\mathbf{r})\rvert^2$ and $\lvert\psi_n^-(\mathbf{r})\rvert^2$, respectively, convoluted with the Fermi function (which is shifted from the Fermi level by $eV$ in presence of a bias voltage $V$). The difference between electron and hole LDOS occurs at the rapid atomic scale oscillation $k_Fr$, because of the phase shift induced by $\eta_n$. This difference is however washed out in the experiment because $k_BT\gg(E_{n}-E_{n-1})\approx \frac{\Delta_0^2}{E_F}$ in 2H-NbSe$_2$, 2H-NbSe$_{1.8}$S$_{0.2}$ and in many other superconductors. As a result, the LDOS shows a electron-hole symmetric patterns.

In presence of anisotropic pairing, as in 2H-NbSe$_2$, we can take into account the hexagonal symmetry of the crystalline lattice by using \cite{Hayashi1998}
\begin{equation}
\Delta_a(\mathbf{r}) = c_a \Delta_0 \tanh(r/\xi_V) e^{-i \theta} \cos 6\theta.  
\end{equation}
The sixfold symmetry breaks the rotational symmetry of the isotropic pairing $\Delta(\mathbf{r})$ for $V=0$, as observed in the experiment, but again it leads to axially symmetric solutions.

\paragraph{Including YSR states.} We now consider the effect of magnetic impurities. We locate magnetic impurities at $\mathbf{r}=\mathbf{r}_{p_i}$. The impurity Hamiltonian contains a magnetic ($J_i$) and non-magnetic ($K_i$) part and we write it as
\begin{equation}
H_{\text{imp}} = \sum_i \left(-J_i \hat{s}\cdot \vec{\sigma}+K_i \tau_z \right) \delta(\mathbf{r}-\mathbf{r}_{p_i}),
\end{equation}
where $\tau_z$ is the Pauli matrix in Nambu space. Here the relation between $J_i$ and $K_i$ can be determined by YSR state energy observed in the experiment. Note that the direction of the magnetic moment is specified by a unit vector $\hat{s}$. The eigenvalues and eigenvectors of $\hat{s}\cdot \vec{\sigma}$ are expressed as $\hat{s}\cdot \vec{\sigma} |s\rangle = s |s\rangle$, where $s=\pm 1$ are eigenvalues. For simplicity, we assume that  $J_i=J$ and $K_i=K$.

The perturbed energy eigenvalues and eigenstates, $(E_{n,s},\psi_{n,s})\rightarrow (\tilde{E}_{n,s},\tilde{\psi}_{n,s})$ can be obtained by solving the following equation constructed in the subspace spanned by the relevant nearest-neighbor states,  
\begin{widetext}
\begin{equation}\label{Heff}
\begin{pmatrix}
E_{n-6} & 0 & W_{n-6,n} & 0 & 0\\  
0 & E_{n-1}+V^{s}_{n-1,n-1} & V^{s}_{n-1,n} & V^{s}_{n-1,n+1} & 0\\
W^{*}_{n-6,n} & V^{s*}_{n-1,n} & E_n +V^{s}_{n,n} & V^{s}_{n,n+1} & W_{n,n+6} \\
0 & V^{s*}_{n-1,n+1} & V^{s*}_{n,n+1} & E_{n+1}+V^{s}_{n+1,n+1} & 0\\
0 & 0 & W^{*}_{n,n+6} & 0 & E_{n+6}
\end{pmatrix}
\begin{pmatrix}
c_{n-6,s} \\
c_{n-1,s}\\
c_{n,s} \\
c_{n+1,s} \\
c_{n+6,s} 
\end{pmatrix} = 
\tilde{E}_{n,s} 
\begin{pmatrix}
c_{n-6,s} \\
c_{n-1,s}\\
c_{n,s} \\
c_{n+1,s}\\
c_{n+6,s}
\end{pmatrix}, 
\end{equation}
\end{widetext}
where 
\begin{align}
V^{s}_{n,n'} &= \int d^2 r~ \psi^{\dagger}_{n,s}(\mathbf{r}) H_{\text{imp}} \psi_{n',s}(\mathbf{r}),\\
W_{n,n'} &= \int d^2 r~ \psi^{\dagger}_{n,s}(\mathbf{r}) \Delta_a(\mathbf{r}) \psi_{n',s}(\mathbf{r}),
\end{align}
and the eigenstate has the form 
\begin{equation}
\tilde{\psi}_{n,s} = \sum_{j,s} c_{j,s} \psi_{j,s},
\end{equation}
with the summation index $j\in \{n-1,n,n+1\}$ for isotropic pairing ($c_a=0$, 2H-NbSe$_{1.8}$S$_{0.2}$) and $j\in \{n-6,n-1,n,n+1,n+6\}$ for anisotropic pairing ($c_a\ne0$, 2H-NbSe$_2$).

\begin{figure*}
\includegraphics[width=2\columnwidth]{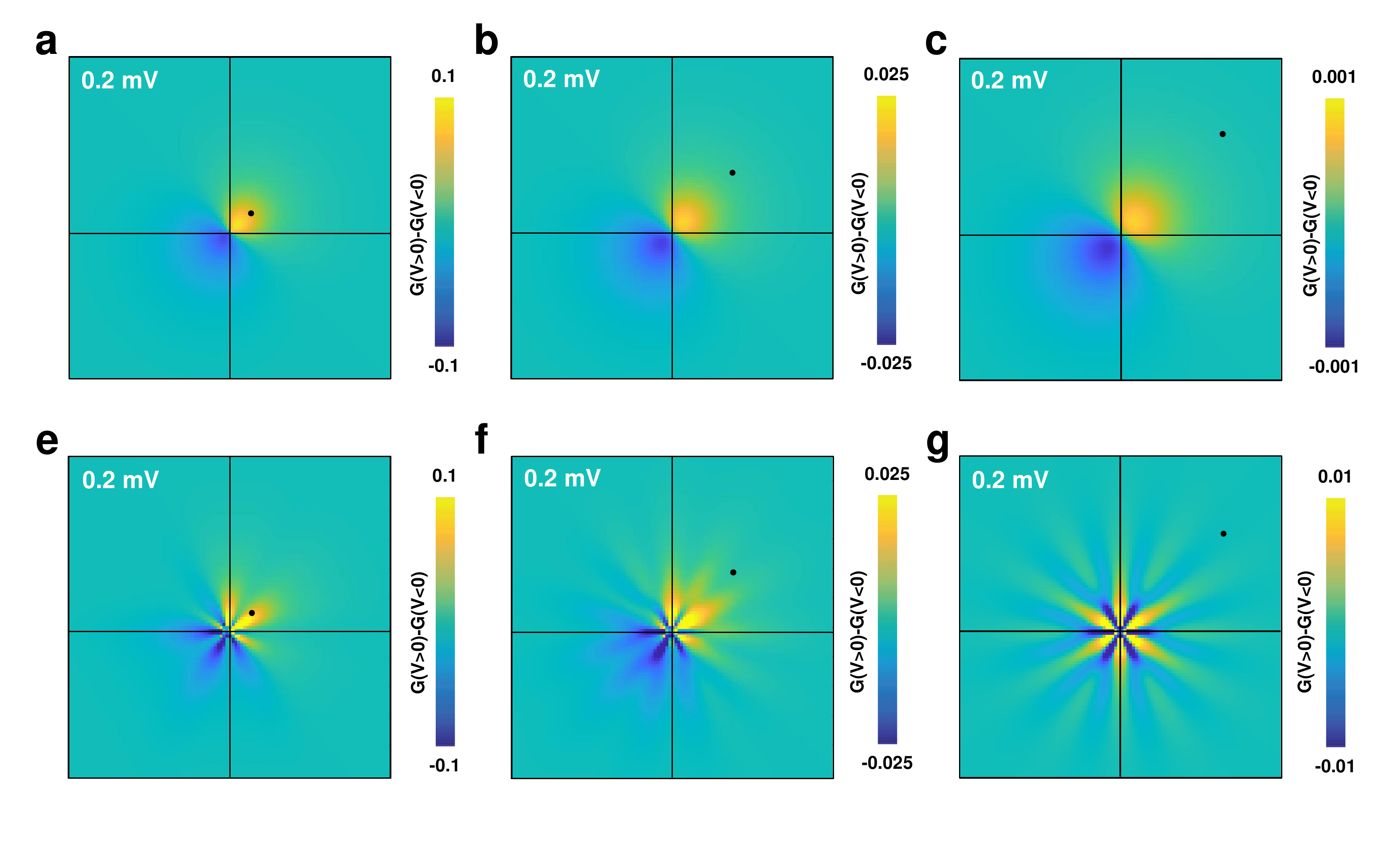}
\caption{{\bf Vortex bound state asymmetry vs. distance of the magnetic impurity from the vortex center.} The calculated difference of the normalized tunneling conductance between positive and negative bias voltages around the vortex center in a field of view of the same size as the ones shown in the main text. The center of the vortex is located at the origin (crossing point between the two black lines) and a single magnetic impurity is marked by a black dot. In {\bf a, e} the impurity is located at 10 nm from the vortex core center and an angle of $\pi/4$, in {\bf b, f} at 30 nm and in {\bf c,g} at 50 nm. Notice the color scale, given by the bars on the right of each figure. The radial asymmetry decreases by several orders of magnitude from {\bf a}-{\bf c} and from {\bf e}-{\bf g}.}
\label{Figure_Supl_Teo}
\end{figure*}

If we neglect the rapid oscillations at the scale of $k_F r$, we can write the probability density difference between the electron-like ($|\tilde{\psi}^{+}_{n,s}|^2$) and the hole-like ($|\tilde{\psi}^{-}_{n,s}|^2$) states as
\begin{equation}
|\tilde{\psi}^{+}_{n,s}(\mathbf{r})|^2 - |\tilde{\psi}^{-}_{n,s}(\mathbf{r})|^2 = \sum_{j,k,s} 
c^{*}_{j,s} c_{k,s} e^{i (j-k) \theta} P_{jk} (r), \label{diffG}
\end{equation}
where 
\begin{align}
P_{jk}(r)=&  A_j(r) A_k(r)  \sin[F_j(r)-F_k(r)] \nonumber\\
 & \times  \sin\left[(\eta_j(r)-\eta_k(r))/2\right]. \label{Pftn}
\end{align}
The difference of the normalized conductance between the positive and negative bias voltages is given by 
\begin{equation}
\frac{\delta G(\mathbf{r},V)}{G_0} = \frac{\beta}{2 \rho} \sum_{n,s} \left[ \tilde{f}_{ns}(V) (|\tilde{\psi}^{+}_{n,s}(\mathbf{r})|^2 - |\tilde{\psi}^{-}_{n,s}(\mathbf{r})|^2)\right],
\end{equation}
where $\beta=1/(k_B T), \rho = m^{*}/(\pi \hbar^2)$ is the normal density of states at the Fermi energy, and 
\begin{equation}
 \tilde{f}_{ns}(V) = \frac{e^{\beta (eV - \tilde{E}_{n,s})}}{\left[1+e^{\beta (eV - \tilde{E}_{n,s})}\right]^2}-
 \frac{e^{\beta (eV + \tilde{E}_{n,s})}}{\left[1+e^{\beta (eV + \tilde{E}_{n,s})}\right]^2}.
\end{equation}

To obtain the results shown in the main text, we use $\xi_V=$30 nm, $k_F$= 9 nm$^{-1}$, $E_F= 135$ meV, $\Delta_0=$ 1 meV, $T=$ 800 mK for both compounds. For 2H-NbSe$_2$, we use $c_a=0.05$, $J= -10$ meVnm$^2$ and $K=10$ meVnm$^2$, whereas for 2H-NbSe$_{1.8}$S$_{0.2}$ we use $c_a=0$, $J= -20$ meVnm$^2$ and $K=20$ meVnm$^2$. $\rho=0.005$ meV$^{-1}$nm$^{-1}$ is used to fit the experimental data. Notice that the position of impurities is very different in both cases (Fig.\,4 of the main text). The different values for $J$ can be associated to difference in the spatial dependence of the wavefunction, which becomes more important when the impurity is close to the vortex center. The actual values of $K$ are not relevant in the calculation of $\frac{\delta G(\mathbf{r},V)}{G_0}$, as we show below. 

To better understand the origin of our result, let us discuss a simple example taking just a single magnetic impurity and isotropic superconducting pairing, $c_a=0$. In the weak perturbation limit, $|c_{n,s}|\gg|c_{n\pm 1,s}|$, the density of the perturbed state can be written as 
\begin{align}
|\tilde{\psi}^{\pm}_{n,s}(\mathbf{r})|^2 \approx&
|c_{n,s}|^2 |\psi^{\pm}_{n,s}(\mathbf{r})|^2 \nonumber\\
& + 2 \text{Re}\left[ c^{*}_{n-1,s}c_{n,s} \psi^{\pm *}_{n-1,s}(\mathbf{r}) \psi^{\pm}_{n,s}(\mathbf{r})\right] \\
& + 2 \text{Re}\left[ c^{*}_{n,s}c_{n+1,s} \psi^{\pm *}_{n,s}(\mathbf{r}) \psi^{\pm}_{n+1,s}(\mathbf{r})\right],\nonumber
\end{align}  
where the coefficients are (up to a normalization factor close to one) 
\begin{align}
c_{n-1,s} &= \frac{V^{s *}_{n,n+1} V^{s}_{n-1,n+1} - V^{s}_{n-1,n} \delta E_{n+1,s}}{D_{n,s}}, \nonumber\\
c_{n,s} &= 1,  \\
c_{n+1,s} &= \frac{V^{s}_{n-1,n} V^{s *}_{n-1,n+1} - V^{s *}_{n,n+1}  \delta E_{n-1,s}}{D_{n,s}},\nonumber
\end{align}
and 
\begin{align}
\delta E_{n+1,s}=& E_{n+1}+V^{s}_{n+1,n+1}-\tilde{E}_{n,s} \nonumber\\
\delta E_{n-1,s}=& E_{n-1}+V^{s}_{n-1,n-1}-\tilde{E}_{n,s} \nonumber\\
D_{n,s} =& \delta E_{n+1,s} \, \delta E_{n-1,s} - |V^{s}_{n-1,n+1}|^2 \label{single-coe}\\
V^{s}_{m,n} =& - s J e^{i (m-n) \theta_p} I_{m,n}(r_p) \nonumber\\
I_{m,n} (r)=& A_{m}(r) A_n(r) \cos[F_{m}(r)-F_{n}(r)] \nonumber\\
& \times\cos[(\eta_{m}(r)-\eta_{n}(r))/2].\nonumber
\end{align}
Using Eq. \eqref{diffG}, we obtain
\begin{align}
|\tilde{\psi}^{+}_{n,s}(\mathbf{r})|^2 - |\tilde{\psi}^{-}_{n,s}(\mathbf{r})|^2 &\approx 
4 J^2 \cos(\theta-\theta_p) P_{n,n+1}(r) \nonumber\\
&\times\frac{I_{n,n+1}(r_p) I_{n-1,n+1}(r_p)}{D_{n,s}}. \label{single-imp}
\end{align}
In the parameter regime we consider, $F_{n}(r)$ and $\eta_{n}(r)$ are monotonically decreasing functions with respect to $n$ and $\eta_{n}-\eta_{n+1}$ is small and positive of the order of $10^{-2}$. In the previous Eq.~\eqref{single-coe}, we ignored the contribution from the non-magnetic potential $K$ in $V^{s}_{m,n}$ because it contains a small factor $\sin((\eta_{m}-\eta_{n})/2)$. The value of $\cos[F_{m}(r)-F_{n}(r)]$ is positive for $0< F_{m}(r) -F_{m+1}(r) < \pi/2$ and negative for $\pi/2< F_{m}(r) -F_{m+2}(r) < \pi$.
The denominator $D_{n,s}$ is in turn negative, so that Eq. \eqref{single-imp} leads to positive $P_{n,n+1}(r)$ and $I_{n,n+1}(r_p) I_{n-1,n+1}(r_p)/D_{n,s}$. So we can write 
\begin{equation}
|\tilde{\psi}^{+}_{n,s}(\mathbf{\theta})|^2 - |\tilde{\psi}^{-}_{n,s}(\mathbf{\theta})|^2 \propto 
J^2 e^{-a r_p/\xi_V}\cos(\theta-\theta_p), \label{single-imp2}
\end{equation} 
where $a\approx 4(1-\varepsilon^2_{n}/2)$. For the case of electron-like bands and within the same simplifying hypothesis one should change $\eta_n(r)$ by $-\eta_n(r)$. It is thus straightforward to conclude that the asymmetry in Eq.(\ref{single-imp2}) would be inverted, i.e. would become $-J^2e^{-a r_p/\xi_V}\cos(\theta-\theta_p)$.

\begin{figure*}
\includegraphics[width=0.8\textwidth]{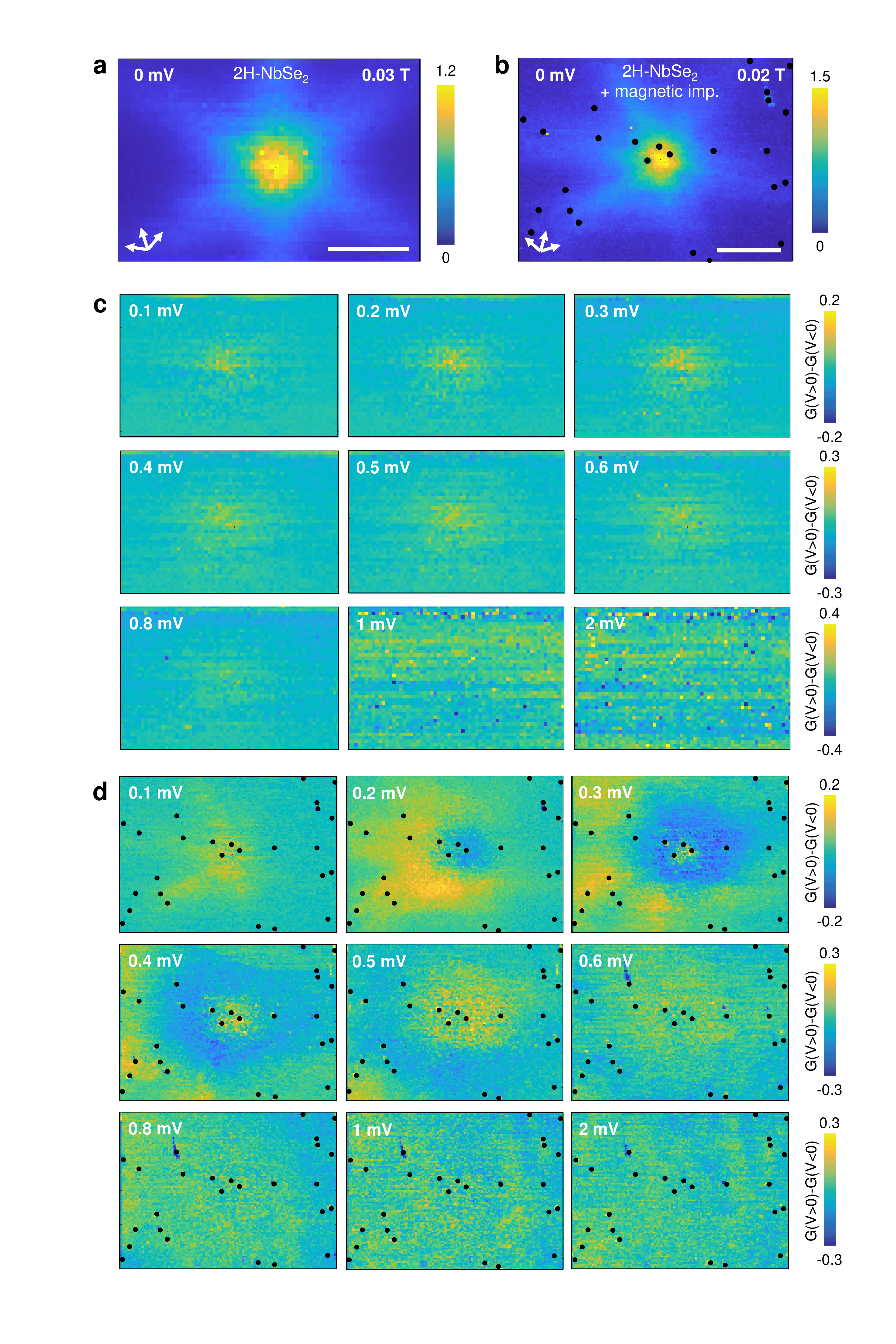}
\vskip -0.5 cm
\caption{{\bf Electron hole symmetry of vortex cores with and without magnetic impurities.} {\bf a} shows a vortex imaged in a field of view without YSR impurities in 2H-NbSe$_2$. We show $\frac{\delta G(\mathbf{r},V)}{G_0}$ of this vortex as a function of the bias voltage (indicated in each panel) in {\bf c}. In {\bf b} we show the same image as in Fig.\,4d of the main text and in {\bf d} we provide $\frac{\delta G(\mathbf{r},V)}{G_0}$ as a function of the bias voltage. White scale bars are 20 nm long. The atomic Se lattice directions are shown by three arrows.}
\label{FigNbSe2_All}
\end{figure*}

\begin{figure}
\includegraphics[width=1\columnwidth]{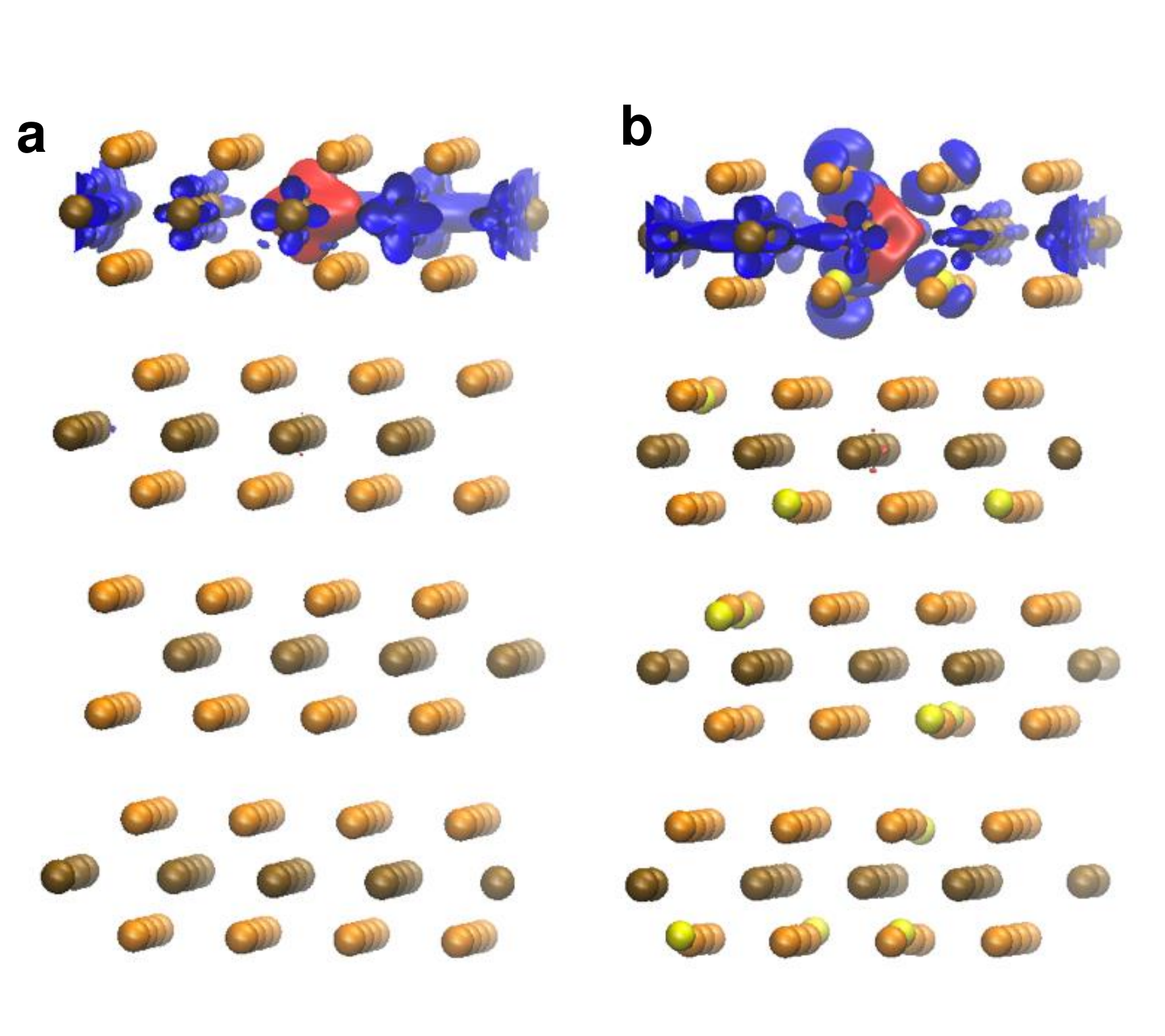}
\caption{{\bf Relaxed atomic positions for 2H-NbSe$_2$ and 2H-NbSe$_{1.8}$S$_{0.2}$ supercells.}  We show the atomic structure of the set of slabs used in the calculation, with Nb atoms in ocre, Se in orange and S in yellow, for supercell A {\bf a} and supercell B {\bf b}. Notice that the atomic positions are slightly modified due to the S substitution in {\bf b}. The distribution of S in {\bf b} is random. We also show a lateral view of the spin polarization due to the substituted Fe atom. We plot the spin isosurface corresponding to a spin imbalance in red (spin up) and blue (spin down).}
\label{Figure_Teo_v2Suppl}
\end{figure}

We show the result of the calculation with a single magnetic impurity in Fig.\,\ref{Figure_Supl_Teo}a-c for $c_a=0$ and d-f for $c_a=0.05$. We use $J=-50$ meVnm$^2$ and $K=50$ meVnm$^2$. We see clearly the angular dependence shown in Eq.\,\eqref{single-imp2}. The impurity induces an electron-hole asymmetry in the CdGM states when it is close to the center of the vortex. The electron-hole asymmetry close to the impurity is compensated by an asymmetry of opposite sign on the other side of the vortex. This breaks the axial symmetry of the vortex LDOS, with a mirror line that joins the vortex center with the impurity. When having many impurities, we add up the effect of each impurity to find the results discussed in the main text. Notice that, when the impurity is far from the vortex core Fig.\,\ref{Figure_Supl_Teo}c, the corresponding asymmetry decreases very rapidly. Therefore, the impurities closest to the vortex cores determine the axial symmetry breaking. 

\begin{figure*}
\includegraphics[width=1\textwidth]{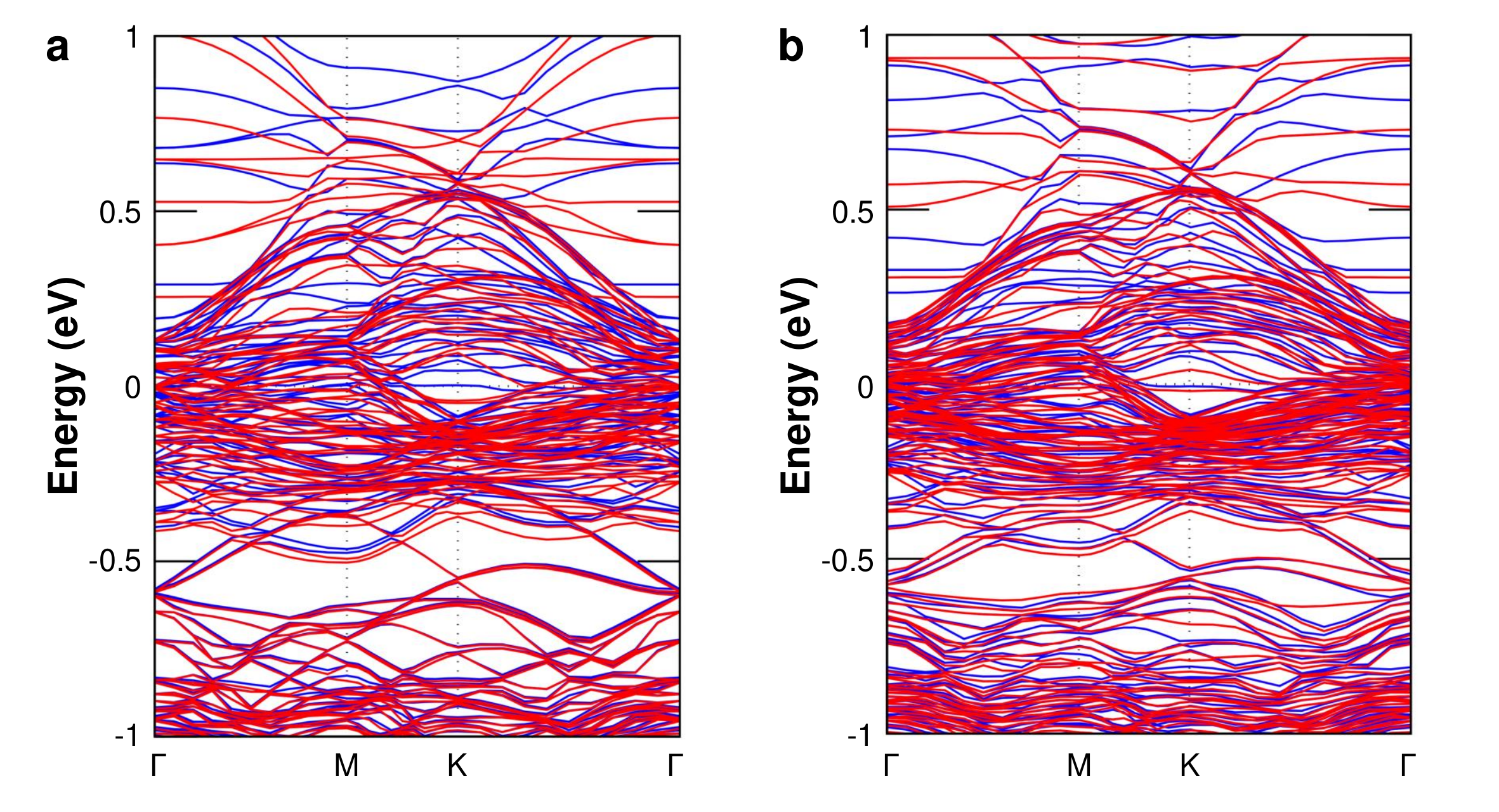}
\caption{{\bf Spin resolved bandstructure of 2H-NbSe$_2$ and of 2H-NbSe$_{1.8}$S$_{0.2}$.} In {\bf a} we show the calculated spin-polarized bandstructure for spin-up (blue) and spin-down (red) states and for 2H-NbSe$_2$ with one Fe atom (supercell A). In {\bf b} we show the same quantities for 2H-NbSe$_{1.8}$S$_{0.2}$ with one Fe atom (supercell B).}
\label{FigSuplDFT}
\end{figure*}

\section{S2. STM results with and without YSR impurities.}
In Fig.\,\ref{FigNbSe2_All} we compare the results obtained in a field of view without YSR impurities (Fig.\,\ref{FigNbSe2_All}a,c), with results with YSR impurities (Fig.\,\ref{FigNbSe2_All}b,d). We observe that the vortex is, for all $\frac{\delta G(\mathbf{r},V)}{G_0}$, axially symmetric in absence of YSR impurities (Fig.\,\ref{FigNbSe2_All}c) and axially asymmetric (Fig.\,\ref{FigNbSe2_All}d) in presence of YSR impurities.

Notice the small electron-hole asymmetry in absence of YSR impurities (Fig.\,\ref{FigNbSe2_All}a,c), which we discuss below (Fig.\,\ref{FigHat}).

\section{S3. Computational details of the spin-polarized electronic bandstructure calculations}

\paragraph{Calculations.} We performed spin-polarized first principles calculations based on DFT with the generalized gradient approximation (GGA) of Perdew-Burke-Ernzerhof\cite{PhysRevLett.77.3865} for the exchange-correlation functional. The plane wave basis sets used projector augmented wave (PAW) pseudopotentials\cite{PhysRevB.59.1758} and the electronic wave functions were expanded with well-converged kinetic energy cutoffs of 75 Ry and 500 Ry for the wavefunctions and charge density, respectively. Dispersion interactions to account for van der Waals interactions between the layers were considered by applying semi-empirical Grimme D2 corrections\cite{doi:10.1002/jcc.20495}.

\paragraph{Relaxed atomic arrangements.} To model the experimental system, we constructed slabs of 4$\times$4$\times$2 size formed by four layers (192 atoms each). The relaxed atomic positions are represented in Fig.\,\ref{Figure_Teo_v2Suppl}, together with a lateral view of the spin isosurfaces discussed in the main text. All the structures were fully optimized without constraints until the forces on each atom were smaller than 10$^3$ Ry/au and the energy difference between two consecutive relaxation steps less than 10$^4$ Ry. The Brillouin zone was sampled by a $\Gamma$ centered 3 $\times$ 3 $\times$ 1 k-point Monkhorst-Pack \cite{PhysRevB.13.5188} mesh for structural optimization and 6 $\times$ 6 $\times$ 2 for the self-consistent field (SCF) calculations. We built two supercells. The supercell A is formed by a single Fe impurity substituting a Nb atom and 63 Nb and 128 Se atoms. The supercell B includes a 10\% random substitution of Se by S atoms, resulting in 1 Fe, 63 Nb, 115 Se and 13 S atoms. 

\paragraph{Magnetic interactions.} We inset a vacuum of 18 \AA\ in between sets of cells, to avoid interaction between replica images as a result of periodic boundary conditions. In order to describe the strong correlation of electrons in Mott-Hubbard physics, we adopted a DFT+U approach, where U is the on-site Coulomb repulsion, using the simplified version proposed by Dudarev et al\cite{PhysRevB.57.1505}. The Hubbard U was estimated self-consistently using density functional perturbation theory\cite{PhysRevB.98.085127}. The distance with the nearest periodic image is ~1.4 nm, which is enough to discard any kind of interaction between the Fe ions. Generally, Nb atoms carry only small magnetic moments, that are oppositely oriented to the Fe moments. On the lateral edges of the supercell, however, we observe a polarization of Nb atoms which is small and might be influenced by the neighboring Fe sites.  All calculations were carried out in the QuantumEspresso code\cite{Giannozzi_2009}.

\begin{figure*}
\includegraphics[width=0.9\textwidth]{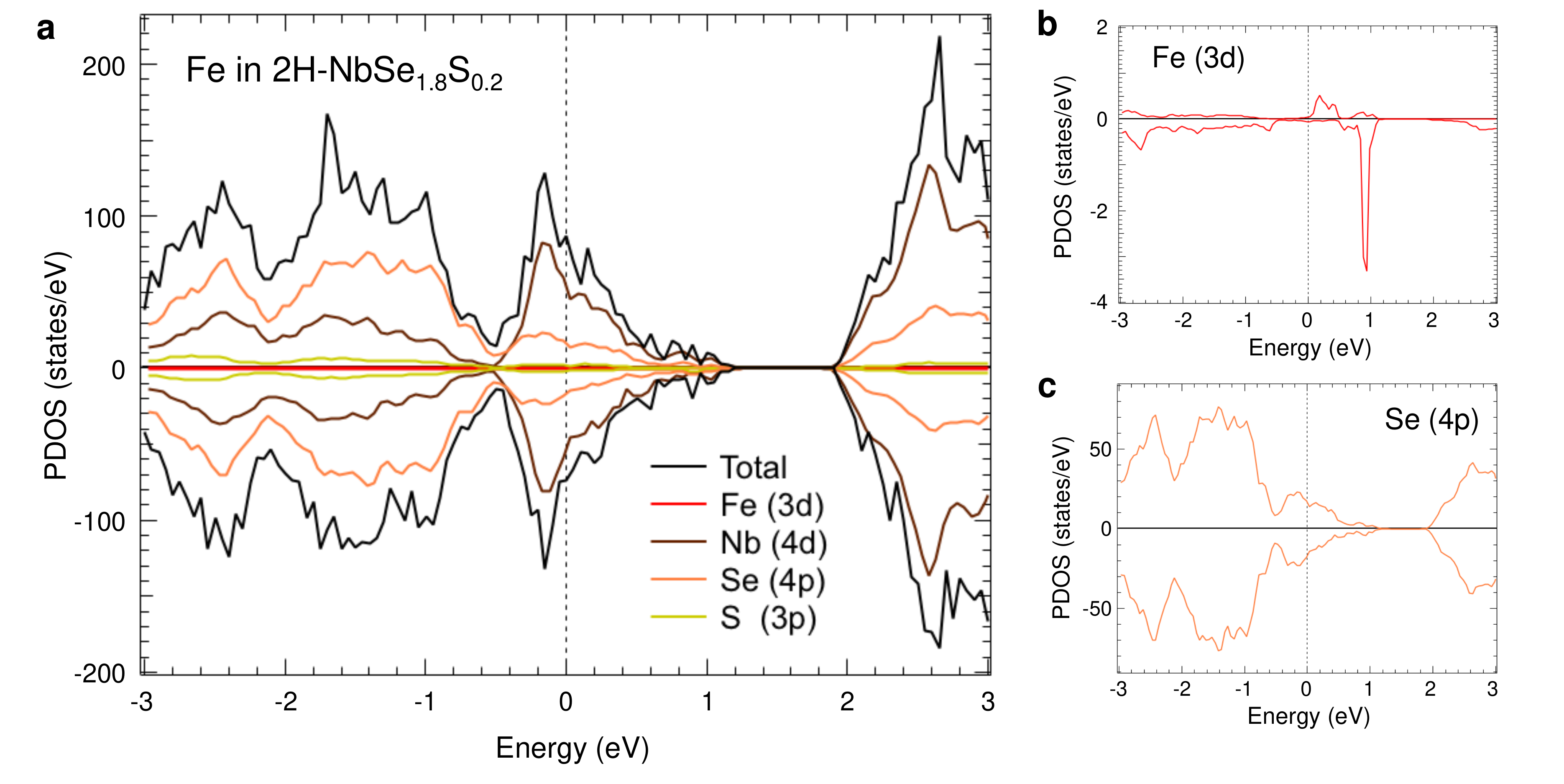}
\caption{{\bf Partial densities of states of atomic species in 2H-NbSe$_{1.8}$S$_{0.2}$.} {\bf a} Spin-polarized partial densities of states integrated over the Brillouin zone for the orbitals marked in the figure for 2H-NbSe$_{1.8}$S$_{0.2}$ with one Fe atom (supercell B). The result for the Fe and Se orbitals is shown in {\bf b} and {\bf c} respectively.}
\label{Figpdos}
\end{figure*}

\paragraph{Bandstructure.}  In the Fig.\,\ref{FigSuplDFT}a,b we plot the resulting spin resolved bandstructures of 2H-NbSe$_2$ and of 2H-NbSe$_{1.8}$S$_{0.2}$ over the whole supercells. We see that the overall energy dependence of the density of states integrated over the whole Brillouin zone is very similar for both cases. The orbital dependent partial densities of states of orbitals that are close to the Fermi level (Se-3p, Nb-4d and S-3p), show nearly the same values for both spin orientations (Fig.\,\ref{Figpdos}), although there are slight but visible differences in the Nb and Se partial densities of states, corresponding to the spin polarization of the atoms located close to the Fe impurity. Of course, the result on the atomic Fe-3d orbitals show a clear spin polarization (inset of Fig.\,\ref{Figpdos}).

\begin{figure}[hbt!]
\includegraphics[width=1\columnwidth]{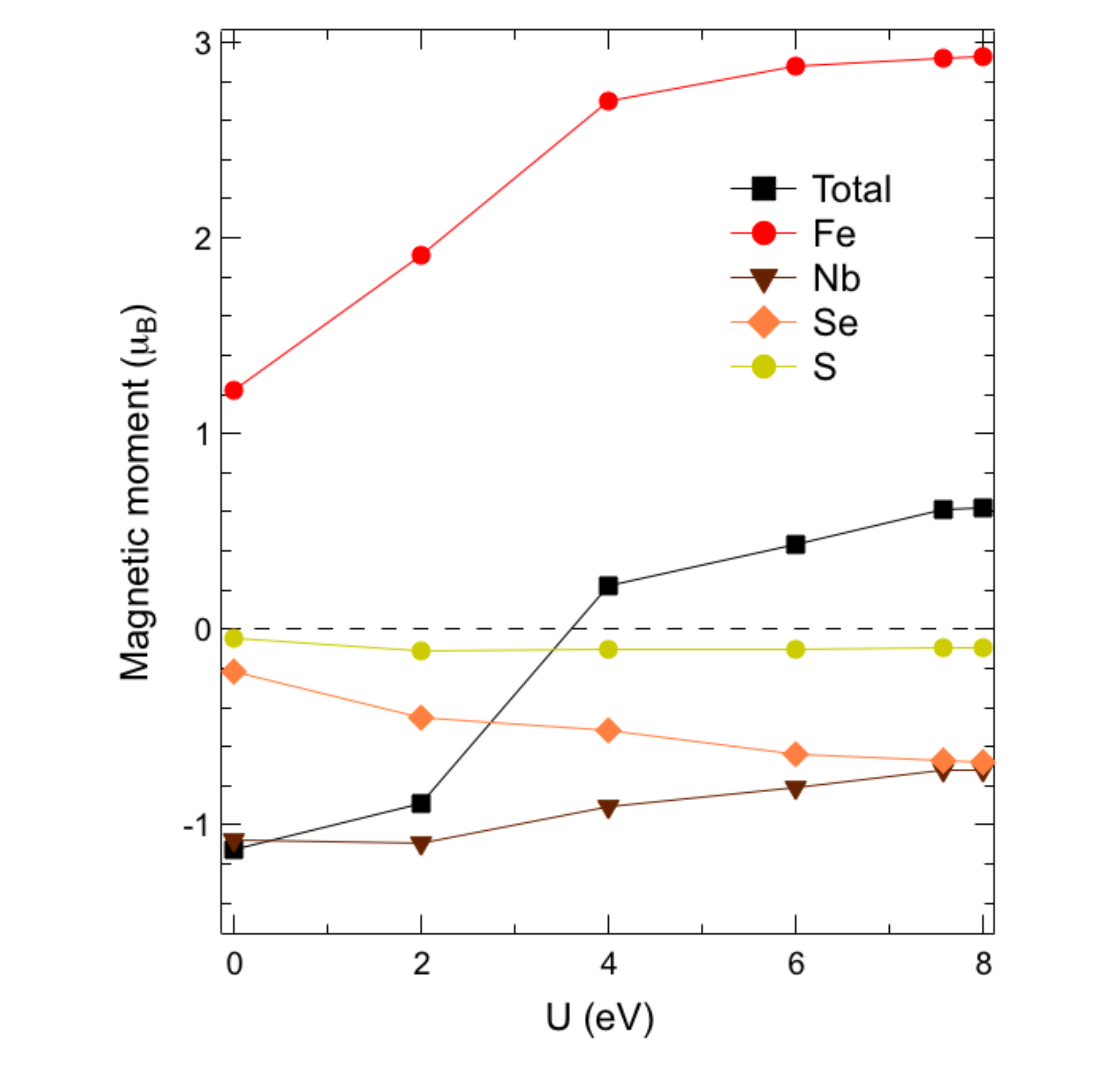}
\caption{{\bf U parameter in calculations.} Magnetic moment as a function of the U parameter for 2H-NbSe$_{1.8}$S$_{0.2}$ with one Fe atom (supercell B).}
\label{FigTeoU}
\end{figure}

\paragraph{Magnetic moment.}  In the Fig.\,\ref{FigTeoU} we show the dependence of the induced magnetic moment with U. We use a reduced system formed by a 4 $\times$ 4 slab of monolayer 2H-NbSe$_2$ or of 2H-NbSe$_{1.8}$S$_{0.2}$ containing one Fe impurity that substitutes one Nb. We see that we reach convergence above about 7eV. We use U = 7.5681 eV in our calculations. However, it is also relevant to remark that the magnetic moment remains large already at relatively small values of U.

\section{S4. STM measurements}

To perform the STM measurements we prepared a plate like sample and glued it to our sample holder. We glued a piece of alumina on top of the sample and removed it at 4.2 K by pushing the piece with a beam. To this end, we used the movable sample holder described in Ref.\cite{Suderow2011}. We measured the freshly exposed surface in a cryogenic system with a base temperature of 800 mK. The design of the STM microscope is very similar to the one described in Ref.\cite{Suderow2011,Galvis2015}. We usually work with tunneling conductances of order of 0.1 $\mu$S or below. We provide the tunneling conductance normalized to its value well above the gap edge, usually between 4 mV and 10 mV. Magnetic fields are applied perpendicular to the plate-like sample.

\section{S5. Crystal synthesis and characterization}

\paragraph{Synthesis.} To synthesize samples of the 2H-NbSe$_2$ and of 2H-NbSe$_{1.8}$S$_{0.2}$ we first mixed powders of Nb, Se (99.999\% Se from Alfa-Aesar) and S (99.98\% S from Sigma-Aldrich) in a stoichiometric ratio, and sealed these in an evacuated quartz ampoule. We heated from room temperature to 900 $^{\circ}$C at 1.5 $^{\circ}$C/min. Then, the temperature was kept constant for ten days and the furnace was switched off for cooling. We mixed 4 mmol of the previously synthesized material with iodine as transport agent (iodine concentration of 5 mg/cm$^3$). We sealed the mixture in an evacuated quartz ampoule and placed it inside a three-zone furnace with the compound in the leftmost zone. The other two zones were heated up in 3 h from room temperature to 800 $^{\circ}$C and kept at this temperature for two days. After that we established a gradient of 800 $^{\circ}$C $/$ 750$^{\circ}$ $/$ 775 $^{\circ}$C in the three-zone furnace. The temperatures were kept constant for 15 days and the furnace then cooled down naturally.

\begin{figure*}
	\includegraphics[width=0.95\textwidth]{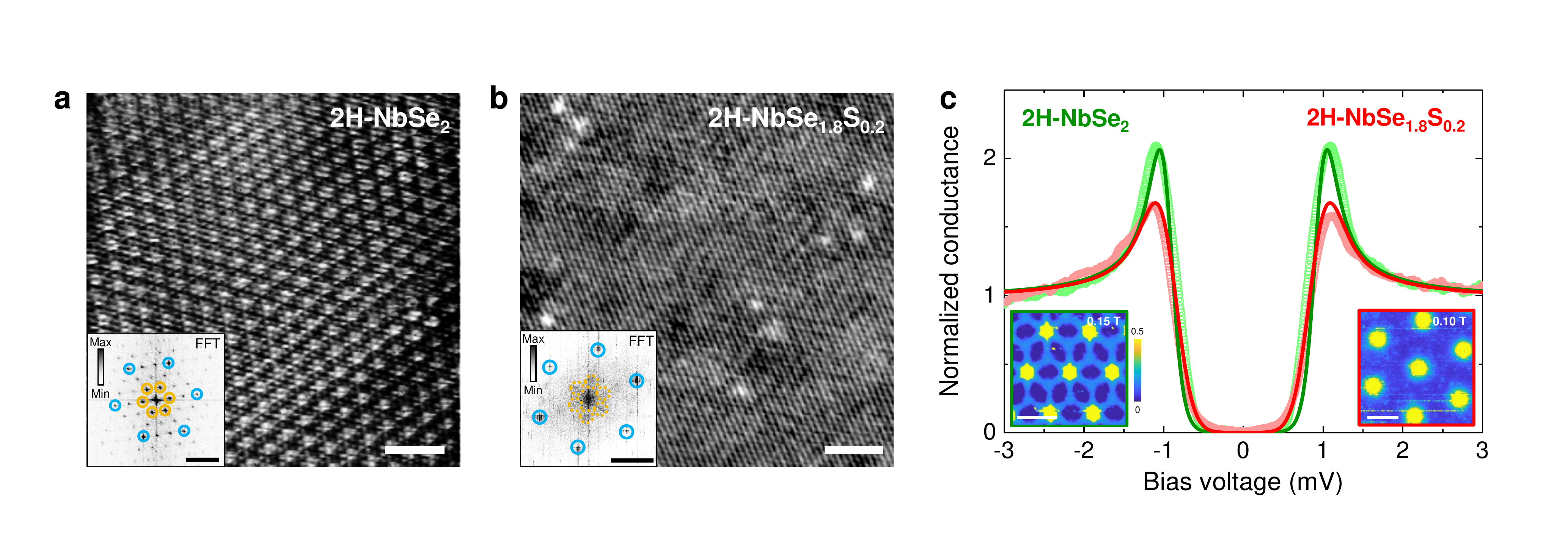}
	\caption{{\bf Topography, superconducting gap and vortex lattice in 2H-NbSe$_2$ and 2H-NbSe$_{1.8}$S$_{0.2}$.} {\bf a} Topographic image of pure 2H-NbSe$_2$. In the bottom left inset we show the Fourier transform of the topography. Atomic Bragg spots are marked with blue circles and CDW Bragg spots with orange circles. {\bf b} Similar figure in a field of view of the same size in 2H-NbSe$_{1.8}$S$_{0.2}$. Scales bars on the bottom right are 3 nm long. {\bf c} Normalized tunneling conductance vs bias voltage in 2H-NbSe$_2$ (light green line, T=0.1 K) and in 2H-NbSe$_{1.8}$S$_{0.2}$ (orange line, T=0.8 K). These data are taken at zero field. The zero bias conductance maps showing the vortex lattice under magnetic fields are given in the lower left and right insets, with the color scale given in the lower left inset. Scale bars in the insets are 100 nm long.}
	\label{Fig2HCompounds}
	\end{figure*}

\begin{figure*}
\includegraphics[width=0.9\textwidth]{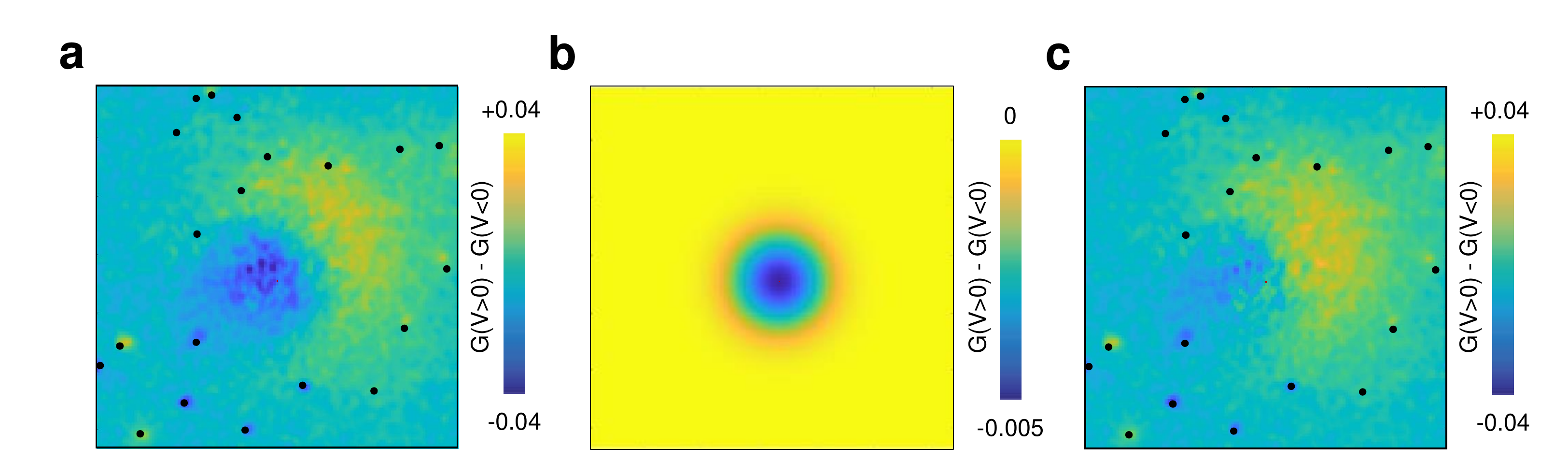}
\caption{{\bf Axially symmetric electron-hole asymmetry.} {\bf a} Difference of the normalized tunneling conductance $G$ between positive and negative bias voltages, $\frac{\delta G(\mathbf{r},V)}{G_0}$ at 0.2 mV is shown as a color scale (bar on the right). {\bf b} Slight electron-hole asymmetric background modelled using a simple gaussian function. {\bf c} Map resulting from the subtraction of {\bf b} from {\bf a}. Black dots in {\bf a} and {\bf c} provide the positions of impurities.}
\label{FigHat}
\end{figure*}

\begin{figure*}
\includegraphics[width=0.9\textwidth]{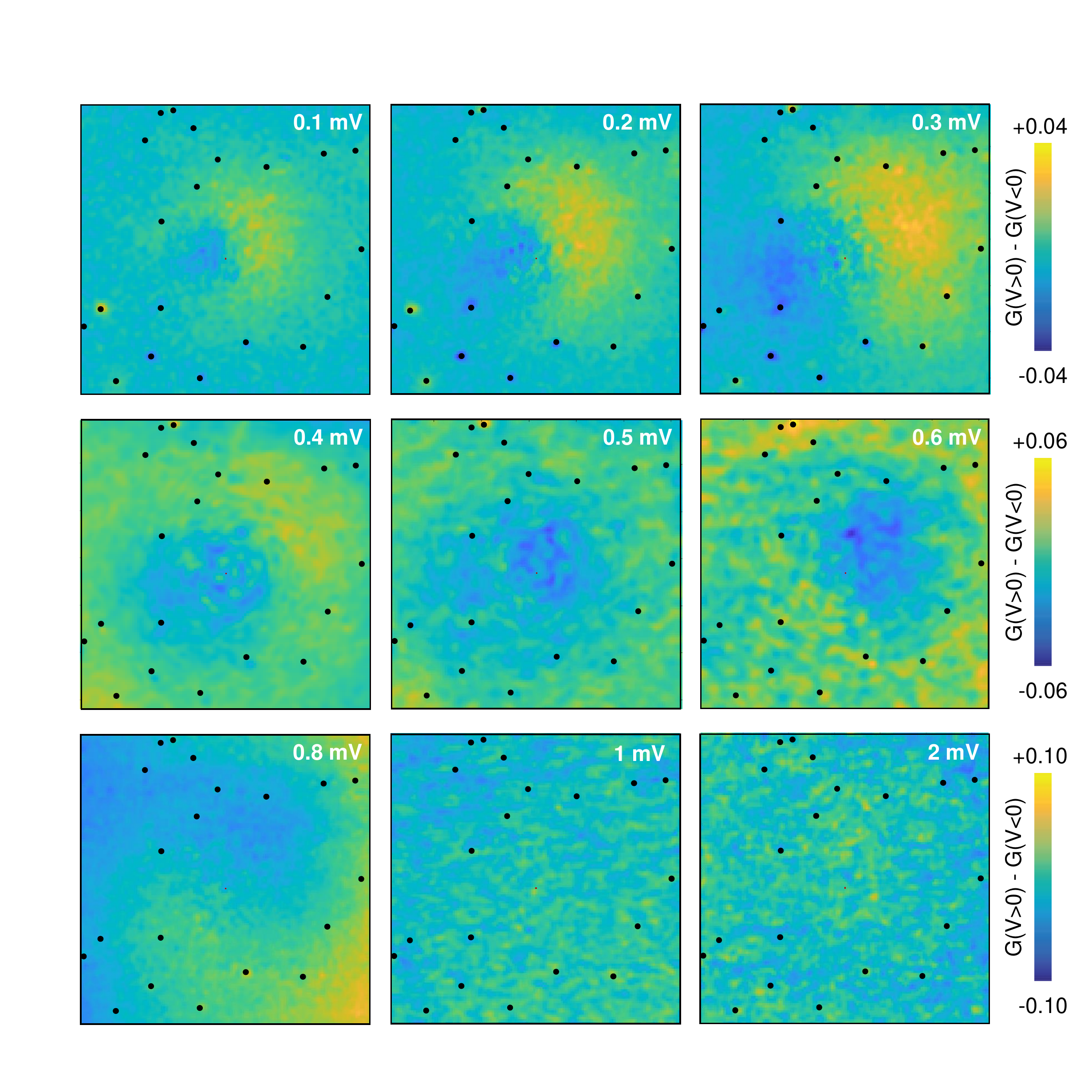}
\caption{{\bf Bias voltage dependence of $\frac{\delta G(\mathbf{r},V)}{G_0}$ in 2H-NbSe$_{1.8}$S$_{0.2}$}. We show in each panel $\frac{\delta G(\mathbf{r},V)}{G_0}=\frac{G(\mathbf{r},V)-G(\mathbf{r},-V)}{G_0}$ for the bias voltages marked in each panel. The field of view is the same as in Fig.\,4 of the main text. Black dots provide the position of magnetic impurities. Color scale is given by the bars on the right.}
\label{FigNbSeS_All}
\end{figure*}

\paragraph{Characterization.} We obtained large single crystals, with lateral sizes in the order of several millimeters. The crystals were analyzed by powder X-ray diffraction and inductively coupled plasma (ICP) mass spectrometry. The experimental powder patterns were refined with the structure of pure 2H-NbSe$_2$ (ICSD 51589) in both 2H-NbSe$_2$ and of 2H-NbSe$_{1.8}$S$_{0.2}$. We obtained a variation in lattice constants with S doping compatible with literature ($a=3.4451(3)$\AA, $c=12.542(1)$\AA\, in 2H-NbSe$_2$ and $a=3.4327(5)$, $c=12.506(2)$ in 2H-NbSe$_{1.8}$S$_{0.2}$) \cite{doi:10.1002/zaac.200500233,PhysRevB.6.835}. Both $a$ and $c$ parameters of the hexagonal structure decrease by the same ratio, a result that remains when increasing the S concentration\cite{SamuelToBePublished}. To ascertain that S doping does not introduce defects other than substution, we have estimated the stacking fault density along the c-axis from the X-ray data. Being a van der Waals compound with little coupling between hexagonal 2H-NbSe$_{2-x}$S$_{x}$ planes, we can expect most defects to occur along the c-axis. Following Ref.\,\cite{warren1990x}, we can distinguish between deformation and growth faults, with respectively probabilities $\mu$ and $\nu$. Deformation and growth faults can be estimated by analyzing reflections of the type H-K=3N$\pm$1. We can then write for the full width at half maximum intensity of the powder scattering Bragg peaks with Miller indices HKL $B_{2\theta}$, $B_{2\theta}=\frac{360}{\pi^2}\tan{\theta}\vert L \vert \left(\frac{d}{c}\right)^2(3\mu+3\nu)$ for even L and $B_{2\theta}=\frac{360}{\pi^2}\tan{\theta}\vert L \vert \left(\frac{d}{c}\right)^2(3\mu+\nu)$ for odd L, with the HKL spacing and $c=2d_{002}$\cite{warren1990x}. We find that the amount of defects along the c-axis is around $2\%$ in both 2H-NbSe$_2$ and 2H-NbSe$_{1.8}$S$_{0.2}$. From inductively coupled plasma (ICP) analysis, we observe 150 ppm of Fe. We do not detect further transition metal impurities within the detection limits of ICP.


The superconducting density of states at zero field of 2H-NbSe$_{1.8}$S$_{0.2}$ shows a smooth distribution of gap values\cite{Fente2016}. It is useful to compare the effect of S substitution with the application of pressure in 2H-NbSe$_2$\cite{PhysRevLett.95.117006,PhysRevResearch.2.043392}. Pressure increases T$_c$ up to 8.5 K at 10GPa, and then T$_c$ is slightly reduced to 7.5 K at 20 GPa. S substitution by contrast decreases T$_c$\cite{Cho2018}. As shown in Ref.\,\cite{Fente2016}, the charge density wave (CDW) of 2H-NbSe$_2$, becomes strongly affected by S substitution in 2H-NbSe$_{1.8}$S$_{0.2}$. In Fig.\,\ref{Fig2HCompounds}a,b we compare topographic STM images of 2H-NbSe$_2$ with 2H-NbSe$_{1.8}$S$_{0.2}$. 2H-NbSe$_2$ shows CDW order with periodic modulations three times the in-plane lattice constant. In 2H-NbSe$_{1.8}$S$_{0.2}$ we also find CDW order at the same wavevector than for $x=0$. However, the intensity of the CDW modulation strongly varies with position, producing a disordered CDW pattern. Thus, the S substitution in 2H-NbSe$_{1.8}$S$_{0.2}$ leads to a superconductor which is very similar to 2H-NbSe$_2$, but in-plane isotropic. The mean free path estimated from the residual resistivity is of about $\ell\approx$ 20 nm in 2H-NbSe$_{1.8}$S$_{0.2}$, significantly below $\ell\approx$ 120 nm in 2H-NbSe$_2$. However, CdGM bound states are well identified in the LDOS\cite{Fente2016}.

\paragraph{Superconducting gap and vortex lattice.} In Fig.\,\ref{Fig2HCompounds}c we show the superconducting gap and vortex lattice in pure 2H-NbSe$_2$ and in 2H-NbSe$_{1.8}$S$_{0.2}$. The results in 2H-NbSe$_2$ have been obtained repeatedly in the past (see e.g. \cite{Hess1989,Hess1990,Guillamon2008c}) and correspond to a superconductor having different values of the gap over the Fermi surface. This is somewhat different in 2H-NbSe$_{1.8}$S$_{0.2}$, which shows a more homogeneous gap distribution. The vortex lattice of 2H-NbSe$_{1.8}$S$_{0.2}$ loses the sixfold star shape characteristic of 2H-NbSe$_2$ and vortices have instead a round shape (bottom insets of Fig.\,\ref{Fig2HCompounds}c)\cite{Fente2016}.

\section{S6. Complete bias voltage dependence in 2H-NbSe$_{1.8}$S$_{0.2}$.}

We have subtracted a radially symmetric signal to $\frac{\delta G(\mathbf{r},V)}{G_0}$ to obtain the images shown in the main text for 2H-NbSe$_{1.8}$S$_{0.2}$. As we show in Fig.\,\ref{FigHat}, the radially symmetric electron hole anisotropy in $\frac{\delta G(\mathbf{r},V)}{G_0}$ is very small, of less than 10\% of $\frac{\delta G(\mathbf{r},V)}{G_0}$. For completeness, we provide the results for all bias voltages in 2H-NbSe$_{1.8}$S$_{0.2}$ in Fig.\,\ref{FigNbSeS_All}. We observe that for bias voltages above 0.3 mV, the in-plane asymmetry is washed out and there is no signal for bias voltages above the superconducting gap.



\section{S7. Magnetic susceptibility measurements and large size conductance maps at zero field and under magnetic fields in the normal phase}

\begin{figure*}
\includegraphics[width=0.9\textwidth]{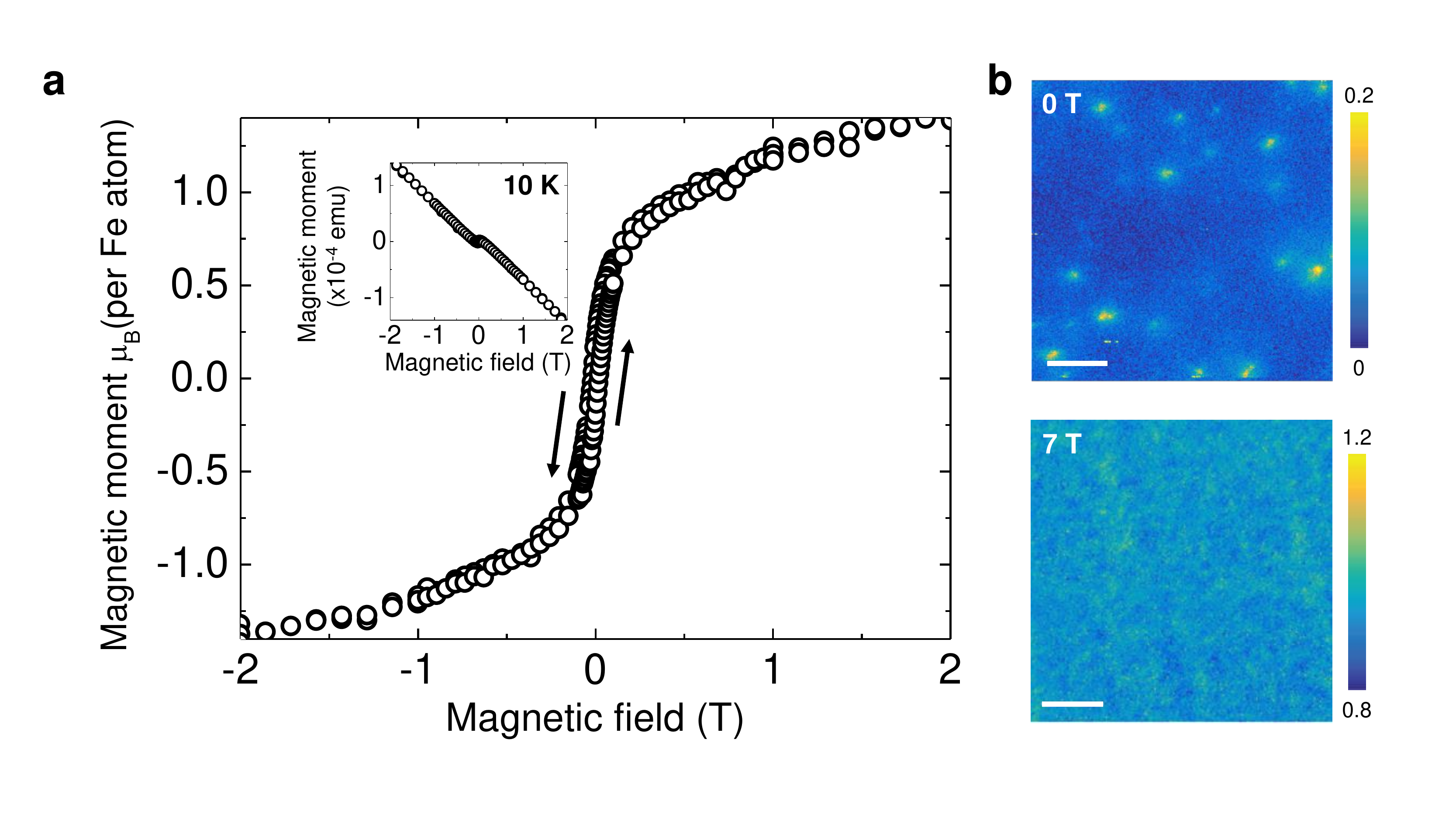}
\caption{{\bf Macroscopic magnetic susceptibility of Fe impurities in 2H-NbSe$_{1.8}$S$_{0.2}$.} {\bf a} We show as circles the magnetic moment as a function of the magnetic field at 10 K. To obtain this curve, we have subtracted a diamagnetic background from the magnetization as a function of the magnetic field (shown in the inset) and divided by the estimated concentration of Fe atoms in the sample, taking the 150 ppm value determined from inductively coupled plasma analysis. Arrows show the direction of the field sweep. {\bf b} Tunneling conductance at zero bias normalized to the tunneling conductance at large bias at zero field (top panel) and at 7 T (bottom panel). Color scale (bars at the right) show the normalized conductance values.}
\label{FigSusc}
\end{figure*}

\paragraph{Magnetic susceptibility of the bulk.} We have performed susceptibility measurements in the same samples measured by STM using a Quantum Design PPMS system, with the magnetic field applied perpendicular to the plate like sample. Inside the superconducting phase, the signal is dominated by the superconducting diamagnetic response. Above $T_c$ we observe a diamagnetic background and a small signal which is ferromagnetic like (Fig.\,\ref{FigSusc}a). We can extract this small ferromagnetic like component from the background and compare its size with the expected Fe moment, taking a Fe concentration of 150 ppm. It is quite remarkable that, although being clearly a very rough approximation which might be strongly influence by clustering at edges or on large defects induced during growth, we obtain a saturation magnetization with a moment of about a Bohr magneton per Fe ion, compatible with the value found in the theoretical calculations (Fig.\,\ref{FigTeoU}). Previous measurements with much larger (30\%) Fe concentration report values up to five Bohr magnetons\cite{VOORHOEVEVANDENBERG1971167}, which are compatible with a Fe4+ valence. The reduced magnetic moment obtained here points to a reduction of the Fe valence, which might be chemically compensated by Se vacancies. More recently, the substitional exchange of Fe atoms in transition metal dichalcogenide MoS$_2$ has been studied in detail, observing directly the exchange of transition metal atoms and finding similarly spin-polarized electronic bandstructure in small cells of transition metal dichalcogenide layers containing one Fe atom\cite{Fu2020}.

\paragraph{Maps at zero field and in the normal phase.} We provide a large size zero bias conductance map at zero field (Fig.\,\ref{FigSusc}b) and under magnetic fields above the critical field of 2H-NbSe$_{1.8}$S$_{0.2}$ (Fig.\,\ref{FigSusc}c). These show that impurities are generally well separated. Furthermore, impurities do not influence the zero bias density of states of the normal phase above the critical field.

\bibliographystyle{naturemag}
\bibliography{NewBibliography}